%% file: 00-main.tex
\documentclass[acmtog,screen]{./ref/acmart}

\usepackage{tikz}
\usepackage[export]{adjustbox}

\usepackage{colortbl}
\usepackage{siunitx}
\usepackage{float}

\usepackage{xcolor}
\usepackage{array}
\usepackage{stfloats}
\usepackage{multirow}
\usepackage{url}
\usepackage{wrapfig}
\usepackage[ruled,linesnumbered,noend]{algorithm2e} % For 
\usepackage{natbib}
\usepackage[noend]{algpseudocode}

\newcommand{\MI}{M_\text{I}}
\newcommand{\MO}{M_\text{O}}
\newcommand{\MC}{M_\text{C}}

\usepackage{prettyref}
\newrefformat{fig}{Fig.~\ref{#1}}
\newrefformat{par}{Section~\ref{#1}}
\newrefformat{appen}{Appendix~\ref{#1}}
\newrefformat{sec}{Section~\ref{#1}}
\newrefformat{sub}{Section~\ref{#1}}
\newrefformat{table}{Table~\ref{#1}}
\newrefformat{tab}{Table~\ref{#1}}
\newrefformat{ass}{Assumption~\ref{#1}}
\newrefformat{alg}{Algorithm~\ref{#1}}
\newrefformat{def}{Definition~\ref{#1}}
\newrefformat{thm}{Theorem~\ref{#1}}
\newrefformat{cor}{Corollary~\ref{#1}}
\newrefformat{lem}{Lemma~\ref{#1}}
\newrefformat{step}{Step~\ref{#1}}
\newrefformat{ln}{Line~\ref{#1}}
\newrefformat{rem}{Remark~\ref{#1}}
\newrefformat{eq}{Equation~\ref{#1}}
\newrefformat{pb}{Problem~\ref{#1}}
\newrefformat{it}{Item~\ref{#1}}
\newrefformat{te}{Term~\ref{#1}}
\def\Eqref Eq:#1:{\eqref{eq:#1}}
\newrefformat{Eq}{Equation~\Eqref#1:}
\newrefformat{App}{Appendix~\ref#1:}

\begin{document}
\title{Robust and Feature-Preserving Offset Meshing}

\author{Hongyi Cao}
\orcid{}
\email{hyc@hdu.edu.cn}
\affiliation{%
  \institution{Hangzhou Dianzi University}
  \country{China}
}

\author{Gang Xu*, Renshu Gu, Jinlan Xu}
\affiliation{%
  \institution{Hangzhou Dianzi University}
  \streetaddress{2\# Street 1158}
  \city{Hangzhou}
  \state{Zhejiang}
  \postcode{310018}
  \country{China}
}

\author{Xiaoyu Zhang}
\affiliation{%
  \institution{Beijing Institute of Spacecraft System Engineering}
  \city{Beijing}
  \country{China}
}

\author{Timon Rabczuk}
\affiliation{%
  \institution{Institute of Structural Mechanics}
  \city{Bauhaus-Universität Weimar}
  \country{Germany}
}

\author{Yuzhe Luo}
\orcid{0009-0008-5796-6144}
\email{yzluo@zju.edu.cn}
\affiliation{%
  \institution{State Key Laboratory of CAD\&CG, Zhejiang University}
  \country{China}
}
\affiliation{%
  \institution{LightSpeed Studios}
  \country{USA}
}
\author{Xifeng Gao*}
\affiliation{%
 \institution{LightSpeed Studios}
   \city{Bellevue}
  \state{Washington}
  \country{USA}
  \postcode{98004}
 }

\begin{abstract}
We introduce a novel offset meshing approach that can robustly handle a 3D surface mesh with an arbitrary geometry and topology configurations, while nicely capturing the sharp features on the original input for both inward and outward offsets. Compared to the existing approaches focusing on constant-radius offset, to the best of our knowledge, we propose the first-ever solution for mitered offset that can well preserve sharp features.
Our method is designed based on several core principals: 1) explicitly generating the offset vertices and triangles with feature-capturing energy and constraints; 2) prioritizing the generation of the offset geometry before establishing its connectivity, 3) employing exact algorithms in critical pipeline steps for robustness, balancing the use of floating-point computations for efficiency, 4) applying various conservative speed up strategies including early reject non-contributing computations to the final output. Our approach further uniquely supports variable offset distances on input surface elements, offering a wider range practical applications compared to conventional methods. 

We have evaluated our method on a subset of Thinkgi10K, containing models with diverse topological and geometric complexities created by practitioners in various fields.  Our results demonstrate the superiority of our approach over current state-of-the-art methods in terms of element count, feature preservation, and non-uniform offset distances of the resulting offset mesh surfaces, marking a significant advancement in the field.
\end{abstract}

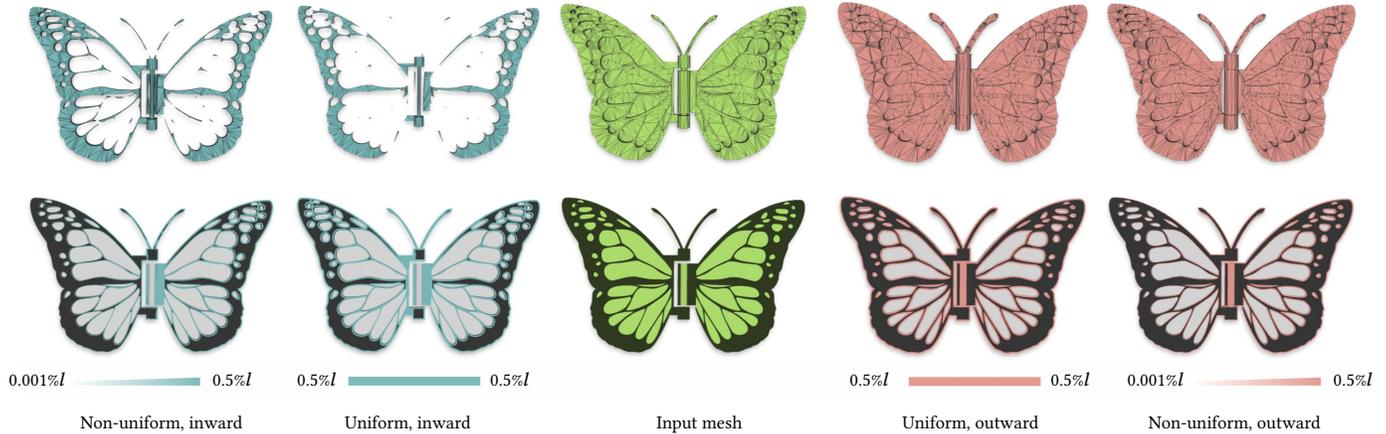
\begin{teaserfigure}
\centering
\includegraphics[width=\linewidth]{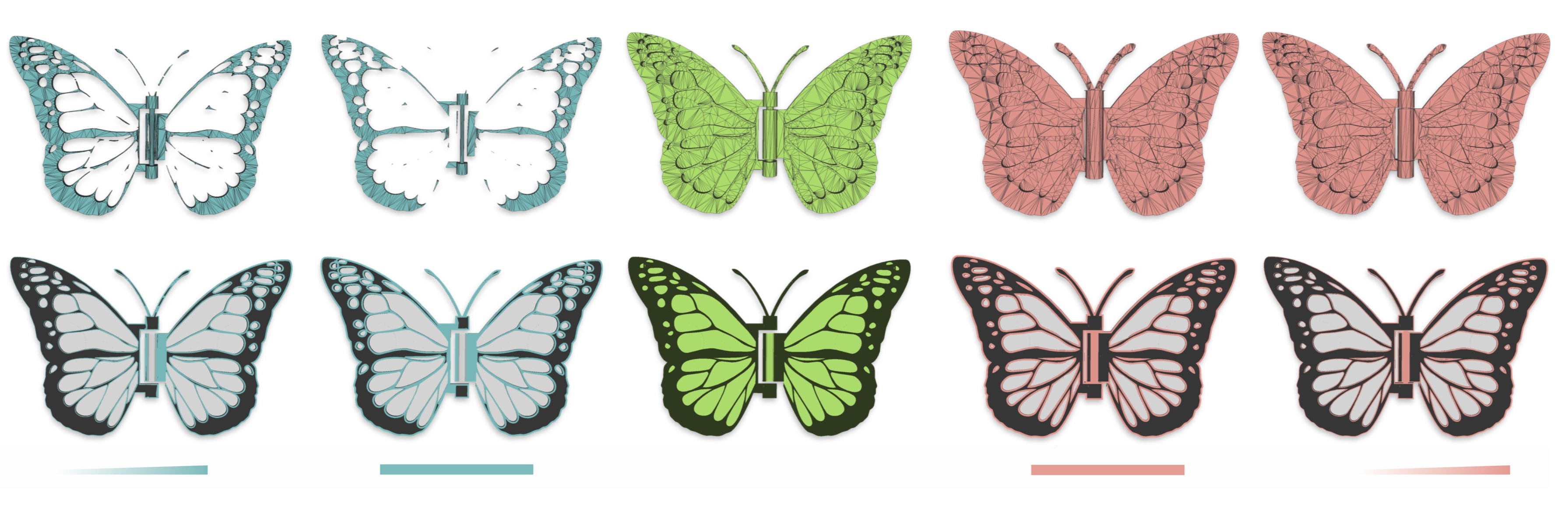}
\put(-515,15){\footnotesize{0.001\%$l$}}
\put(-438,15){\footnotesize{0.5\%$l$}}
\put(-488,-1){\footnotesize{Non-uniform, inward}}
\put(-406,15){\footnotesize{0.5\%$l$}}
\put(-333,15){\footnotesize{0.5\%$l$}}
\put(-388,-1){\footnotesize{Uniform, inward}}
\put(-270,-1){\footnotesize{Input mesh}}
\put(-197,15){\footnotesize{0.5\%$l$}}
\put(-121,15){\footnotesize{0.5\%$l$}}
\put(-177,-1){\footnotesize{Uniform, outward}}
\put(-92,15){\footnotesize{0.001\%$l$}}
\put(-14,15){\footnotesize{0.5\%$l$}}
\put(-84,-1){\footnotesize{Non-uniform, outward}}
\vspace{-0.5em}
\caption{For a 3D mesh with arbitrarily complex geometry and topology, our approach can robustly produce its inward and outward  offset surfaces, according to user desired either uniform or varying offset distances. We assign the non-uniform offset distance to each face of the input by linearly interpolating 0.001\% and 0.5\% of the diagonal length of the input's bounding box. The top row shows the full views of the meshes and the bottom is their corresponding cut views.}
\label{fig:teaser}
\end{teaserfigure}

    


\maketitle

\input{01-intro}
\input{02-related}
\input{03-method}

\input{04-experiments}

\input{05-conclusion}

% \begin{figure*}[t]
% \centering
% \scalebox{1.0}{\includegraphics[width=\linewidth]{fig/gallery_1.jpg}}
% \caption{Result gallery for the entire test dataset using offset distance $d = 0.5\%l$ and $d = 1\%l$. For each model, the left, middle, and right represent the inward, input, and outward meshes, respectively.}
% \label{fig:gallery}
% \end{figure*}

% \begin{figure*}[t]
% \centering
% \scalebox{1.0}{\includegraphics[width=\linewidth]{fig/gallery_2.jpg}}
% \caption{Result gallery for the entire test dataset using offset distance $d = 0.5\%l$ and $d = 1\%l$. For each model, the left, middle, and right represent the inward, input, and outward meshes, respectively.}
% \label{fig:gallery}
% \end{figure*}

\newpage
\bibliographystyle{./ref/ACM-Reference-Format}
\bibliography{06-ref}
\newpage

\clearpage

% \begin{figure}[ht!]
% \centering
% \includegraphics[width=\linewidth]{fig/alphav6.jpg}
% \put(-230,-8){\footnotesize{input mesh}}
% \put(-165,-8){\footnotesize{$\alpha=0.1$}}
% \put(-105,-8){\footnotesize{$\alpha=0.01$}}
% \put(-45,-8){\footnotesize{$\alpha=0.001$}}
% \caption{\label{fig:alpha} A larger value of $\alpha$ leads to smoother offset changes.}
% \end{figure}

\end{document}

%% file: 01-intro.tex
\section{Introduction}\label{sec:intro}
Offset mesh generation, which involves creating a parallel surface with a specific distance from a given shape, holds a place of critical importance in geometric modeling and mesh processing. This technique is fundamental in a variety of applications, including but not limited to computer-aided design and engineering, real-time rendering and animation, robotics, medical imaging, architectural design \cite{PHAM1992223,MAEKAWA1999165}. For example, it is instrumental for designing mechanical parts with specific thickness requirements, such as gears and casings. In the world of animation and 3D modeling, offset surfaces enable the creation of intricate and realistic characters and environments, providing the necessary depth and complexity. 
\setlength{\columnsep}{5pt}
\begin{wrapfigure}{r}{0.6\linewidth}
\flushright
\vspace{-1.0em}
\scalebox{1}{\includegraphics[width=\linewidth]{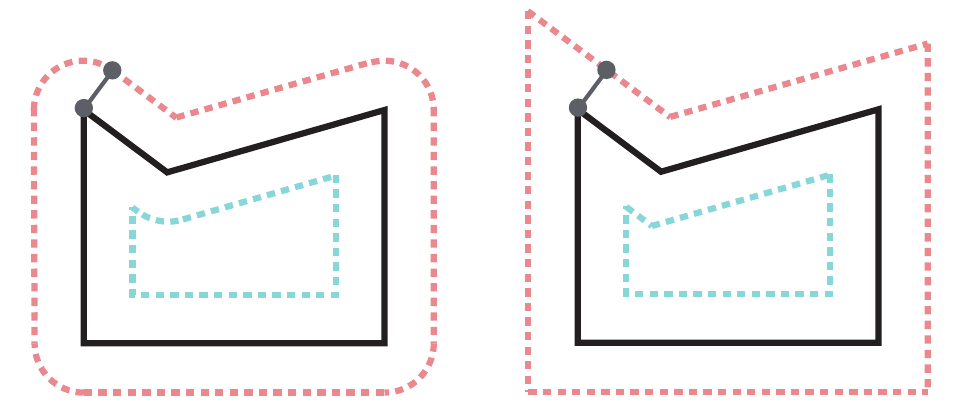}
\put(-135,48){\scriptsize{r}}
\put(-61.2,48){\scriptsize{r}}
    \put(-141,-10){\footnotesize{Constant-radius offset}}
\put(-57,-10){\footnotesize{Mitered offset}}
}
\vspace{-1.0em}
\end{wrapfigure}
Many prior works are proposed to generate offset surfaces for input meshes, such as explicit offsetting following normal directions \cite{jung2004self}, implicit distance function based iso-surfacing \cite{zhen2023,zint2023feature}, Minkowski Sum based boolean approach \cite{hachenberger2009exact}, and offsetting by ray-reps based method \cite{chen2019half}. However, differently from the well-studied curve offset research in 2D \cite{2Doffset84,Clipper1992,park2003mitered} that have robust solutions for both constant radius and mitered offsets (see the inset), to the best of our knowledge, none of existing approaches in 3D can generate offset meshes in the mitered manner. Moreover, the methods for even constant radius offset typically suffer one or more of the following critical issues that hinder their practical usage: 1) lack of robustness in dealing with ``dirty'' data that may have open boundaries, non-manifold vertices and edges, and self-intersections, which is common in scanned or man-made models; 2) struggle to maintain the fidelity of the original shape, especially in handling complex geometries with sharp edges and intricate details, resulting in undesired artifacts in the offset surface; 3) computational inefficiency when the offset distance is small, which can often jointly pose memory issues. %As an example, \prettyref{fig:issues} illustrates some of these issues when performing the current state-of-the-art approaches on a sample 3D model. 
These issues collectively underscore the need for an innovative solution that can address the complexities of modern geometric shapes while ensuring computational efficiency and geometry fidelity.

In this work, we propose a new explicit surface offset generation method that can address the aforementioned issues. The robustness of our approach lies in that we make no assumptions of the input mesh, such as its manifoldness, watertightness, absence of self-intersections and degeneracies, etc, reformulating the offset meshing problem and algorithm design accordingly. We further employ exact algorithmic for robust geometric computations. 
The feature preservation is ensured through first redefining the offset distance from traditionally employed point-to-point to point-to-plane, and then generating the offset surface geometry that can capture sharp features by solving local quadratic energies and dynamic programming. While employing the rational number representation for exact computations, we strive to be efficient by designing several acceleration strategies including parallelization through spatial domain decomposition and dynamic programming for early rejection of the expensive intersection computations.

Our solution has the additional advantages over existing approaches: 1) the resulting offset mesh is similar with the input, from both the number of triangles and connectivity aspects; 2) we support variable offset distances over different surface regions, enabling broader applications than conventional offset methods. 

We empirically compare our approach against state-of-the-arts on a subset of the Thingi10K dataset \cite{Thingi10K}. Our method exhibits a significant improvement in terms of feature preservation, robustness, and element count at the same time. All the data shown in the paper can be found in the attached material, including a video to show their visual quality.

%% file: 02-related.tex
\section{Related Work}
We briefly review prominent methods related to offset meshing, which are based on \textit{direct offsetting}, \textit{distance field}, \textit{Minkowski sum}, \textit{medial axis and skeleton}, and \textit{ray casting}.

\paragraph{Direct Offsetting} A prominent method proposed by \cite{jung2004self} offsets based on vertex normal direction. Although efficient, it struggles with holes in complex CAD models, leading to offset meshes that can be defective, particularly in scenarios involving self-intersections. The precision issues arising from floating-point operations in self-intersections have been addressed using infinite precision operations \cite{campen2010polygonal}, but results occasionally produce twisted meshes. Offsetting surfaces with polynomials or B-splines is explored in \cite{maekawa1999overview}. However, subsequent intersection operations are intricate. Another approach calculates offset positions based on distance and uniform distribution, followed by point cloud reconstruction \cite{meng2018efficiently}.

\paragraph{Distance Field}
These methods are popular in modern 3D printing \cite{brunton2021displaced}.
Generally, they require resampling to generate the final mesh, which can compromise geometric features, especially in detailed meshes \cite{wang2013thickening}.
Challenges also arise from the grid density, preservation of sharp features, and computational efficiency.
Some attempts, such as \cite{kobbelt2001feature}, have tackled the ambiguity of the Dual Contouring and Marching Cube, but they tend to excel mostly with CAD models. In the study by \cite{liu2010fast}, there are some improvements in computational efficiency.
The method introduced in \cite{pavic2008high} tries to preserve sharp features, but it involves high grid density and computational expenses.
An Octree-based method is proposed in the study by \cite{zint2023feature}, which requires the input to be degenerate-free and may output offsets with self-intersections. While the feature-preserving approach introduced in \cite{zhen2023} ensures nice geometry and topology properties of iso-surfaces, it cannot handle offset distances smaller than the grid size which is a common shortcoming shared all distance field based offset extraction methods.

Among the vast offset generation literature, we realize only one work \cite{chen2019variable} tried to address the non-uniform offset problem based on dual-contouring 
\cite{ju2002dual}. However, the approach would require an impractical grid resolution at regions with a small offset distance for tolerable offset reconstruction. More importantly, without much details of how the interpolation is performed for different offset distances within a grid cell, we infer that it either is computationally expensive for proper in/out sign determinations, 
%i.e. $O(N_\mathcal{G}N_\mathcal{T})$ where $N_\mathcal{G}$ is the number of grids and $N_\mathcal{T}$ is the number of triangles of the input,
or has discontinuity issues leading to bumpy artifacts. Moreover, their generated offset meshes would often have self-intersections, even if the input is both topologically simple and geometrically high quality.

\paragraph{Minkowski Sum Method}The Minkowski sum offers a solution for mesh offsetting by calculating the sum of mesh and sphere polygons \cite{rossignac1986offsetting}. An advanced method in \cite{hachenberger2009exact} computes the exact 3D Minkowski sum of non-convex polyhedra by decomposing them into convex parts. Despite its robustness \cite{cgal:eb-23b}, the method is slow for complex CAD models and struggles with variable thickness offsets and sharp feature preservation.

\paragraph{Skeleton-Based Methods}
Skeletal meshes are prevalent in geometry processing \cite{tagliasacchi20163d}. The mesh model can be represented by medial axes and spheres \cite{amenta2001power, sun2015medial}. There are techniques to generate skeleton meshes \cite{lam1992thinning, li2015q}. However, they usually cannot handle general models with open boundaries, self-intersections, etc. There have been efforts to simplify medial axes \cite{sun2013medial}. Offsetting can be achieved by adjusting the radius of the balls on the skeleton. Still, these methods often fall short with  models with sharp features.

\paragraph{Ray-Based Methods}
This approach \cite{chen2019half, wang2013gpu} is rooted in the dexel buffer structure \cite{van1986real}. A recent algorithm in \cite{chen2019half} offers an efficient parallelized method using rays and voxels to compute mesh offsets, bypassing the need for distance field computation. Nonetheless, due to the grid's voxel-like structure, a high density is essential for accurate shape representation, leading to increased computational costs and dense mesh outputs. The approach also mainly considers a consistent offset, further steps have to be designed for the mesh representation conversion.

%% file: 03-method.tex
\section{Method}
In this section, we provide our problem statement, give the pipeline overview, explain each step in detail, and introduce the performance improvement strategies.
\paragraph{Problem Statement:} The input of our approach contains a 3D triangle mesh $\MI$ with an arbitrary geometry and topology configuration, an integer variable $s \in\{-1, 1\}$ to indicate the offsetting direction, and an offset scalar $d_i$ per $T_i$ of $\MI$ where $d_i$ could vary for different triangles. While locally to a triangle, $s = 1$ means the offset direction has a positive product with the triangle's normal and $s = -1$ is for the opposite direction, globally we employ the generalized winding number \cite{jacobson2013robust} for inward and outward checks to robustly handle $\MI$ with possibly gaps, duplicated elements, etc. 

Our goal is to generate an offset mesh $\MO$ that satisfies three qualitative requirements. First, $\MO$ should have the \emph{offset distance} to $\MI$ satisfying user desired value. Second, $\MO$ needs to have nice geometry and topology properties to enable the easy design of automatic downstream algorithms. Third, $\MO$ should contain as few as possible elements such as triangles, allowing efficient computations for the various applications of offset. While hard to properly perform a quantitative measure, we further require $\MO$ to capture the geometry feature of $\MI$ as much as possible, especially for prominent sharp features since we aim for reproducing the mitered offset effect.

\begin{figure}[t]
\centering
\scalebox{1}{\includegraphics[width=\linewidth]{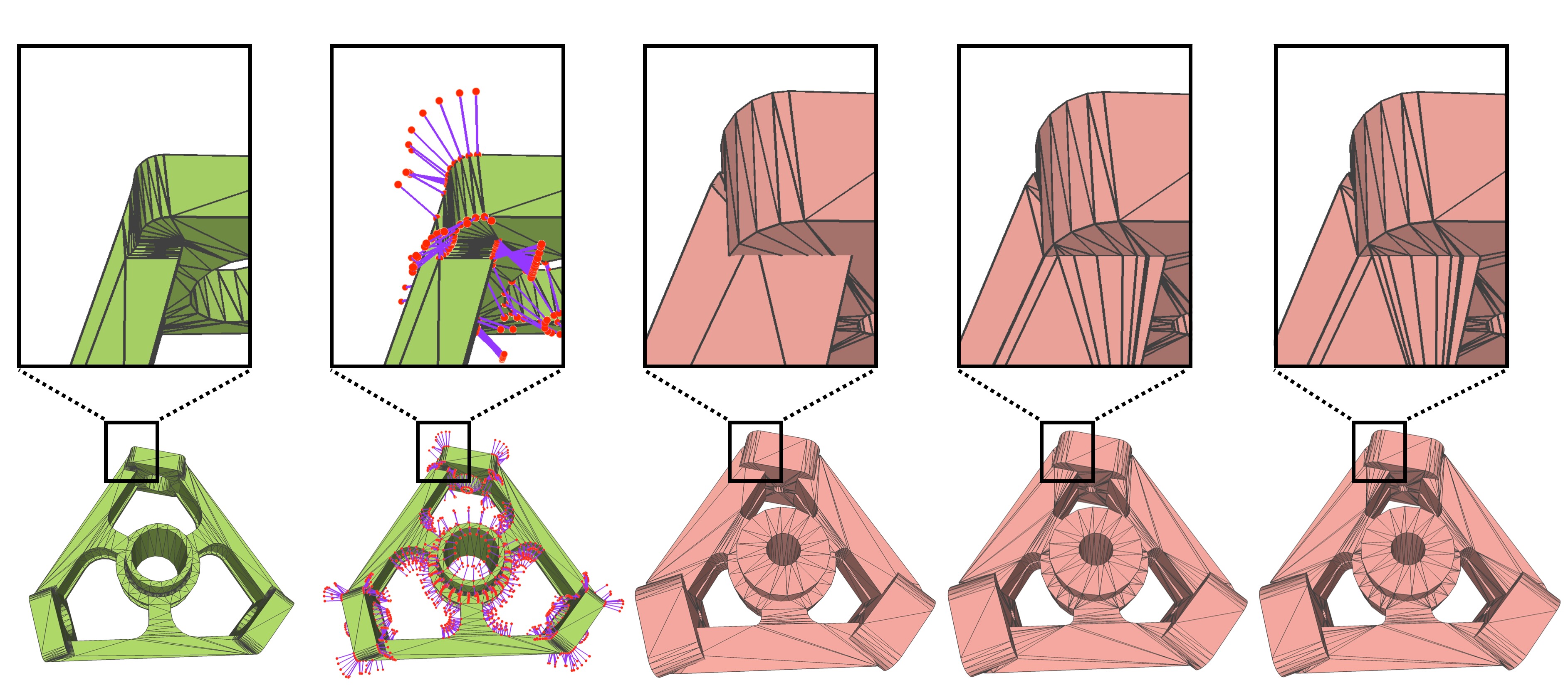}}
% \scriptsize{
% \put(-246,-8){$\MI$ (663, 1326)}
% \put(-205,-8){(1358, 1326)}
% \put(-160,-8){(8365, 10668)}
% \put(-110,-8){$\MC$ (6825, 2275)}
% \put(-54,-8){$\MO$ (788, 1576)}
% }
% }
% \put(-483,-8){\small{$v:$ 663}}
% \put(-483,-17){\small{$f:$ 1326}}
% \put(-383,-8){\small{$v:$ 1358}}
% \put(-383,-17){\small{$f:$ 1326}}
% \put(-275,-8){\small{$v:$ 8365}}
% \put(-275,-17){\small{$f:$ 10668}}
% \put(-165,-8){\small{$v:$ 6825}}
% \put(-165,-17){\small{$f:$ 2275}}
% \put(-50,-8){\small{$v:$ 788}}
% \put(-50,-17){\small{$f:$ 1576}}
\caption{Pipeline : starting from a triangle mesh $\MI$ (left most), our approach first generates its vertices' offset points (second to the left), then builds an offset polyhedron for each of its triangle, edge and vertex (middle). After that, we convert the polyhedra set to an intersection-free triangle soup $\MC$ by filtering out those triangles not part of $\MO$ (second to the right), 
Along with some acceleration measures, this approach involves selecting only a subset of triangles from the polyhedra set for computation, which may result in the creation of some holes. Finally, by detecting the boundaries of the holes, we retrieve the triangles that were excluded in the previous step, Then we construct the connectivity of $\MO$ (right).}
%$(\bullet, \bullet)$ denotes the numbers of vertices and faces of the corresponding mesh, respectively.
\label{fig:pipeline}
\end{figure}
\paragraph{Method Overview:}
As illustrated by \prettyref{fig:pipeline}, we tackle the surface offset problem by first offsetting each vertex of $\MI$ to one or more corresponding points through the solving of a point-to-plane constrained quadratic energy (\prettyref{sub:vo}, \prettyref{fig:pipeline} second to the left), and then constructing a polyhedron representing the local offset volume corresponding to each vertex, edge, and triangle of $\MI$ \prettyref{sub:to}, \prettyref{fig:pipeline} middle). Using a set of acceleration strategies, we then perform boolean operations of the offset volumes to obtain a triangle soup, $\MC$, representing the geometry of $\MO$ (\prettyref{sub:ge}, \prettyref{fig:pipeline} second to the right), which are finally connected to complete the generation of user desired $\MO$ (\prettyref{sub:tc}, \prettyref{fig:pipeline} right).

% \setlength{\columnsep}{10pt}
% \begin{wrapfigure}{r}{0.4\linewidth}
% \vspace{-0.1em}
% \flushright
% \includegraphics[width=\linewidth]{fig/2dsplit.png}
% \vspace{-1em}
% \caption*{\label{fig:split2d} {The diagram illustrates the principle of vertex splitting in a 2D scenario. Black lines depict a local part of the input mesh, and blue dots represent the offset points. We find that in 2D, when the angle between the normal vectors of two line segments is greater than 90 degrees, it is not feasible to use a single point to simultaneously satisfy the corresponding distance constraints. The underlying reason is the contradiction in the respective components upon decomposing the normal vectors of the two line segments. The figure on the far right depicts two overlapping line segments with opposing normal vectors.}}
% \vspace{-1.0em}
% \end{wrapfigure}

\subsection{Vertex Offset}\label{sub:vo}
This step is to generate the offset points for each vertex $V_i\in\MI$. As illustrated in the inset, $V_i$ might correspond to one or more points depending on different local configurations, such as non-manifoldness, saddle region etc. For easy explanation, we first introduce the simple but effective linear constrained quadratic optimization solution when $V_i$ has only one offset point, and then \setlength{\columnsep}{10pt}
\begin{wrapfigure}{r}{0.4\linewidth}
%\vspace{-1.5em}
\flushright
\includegraphics[width=\linewidth]{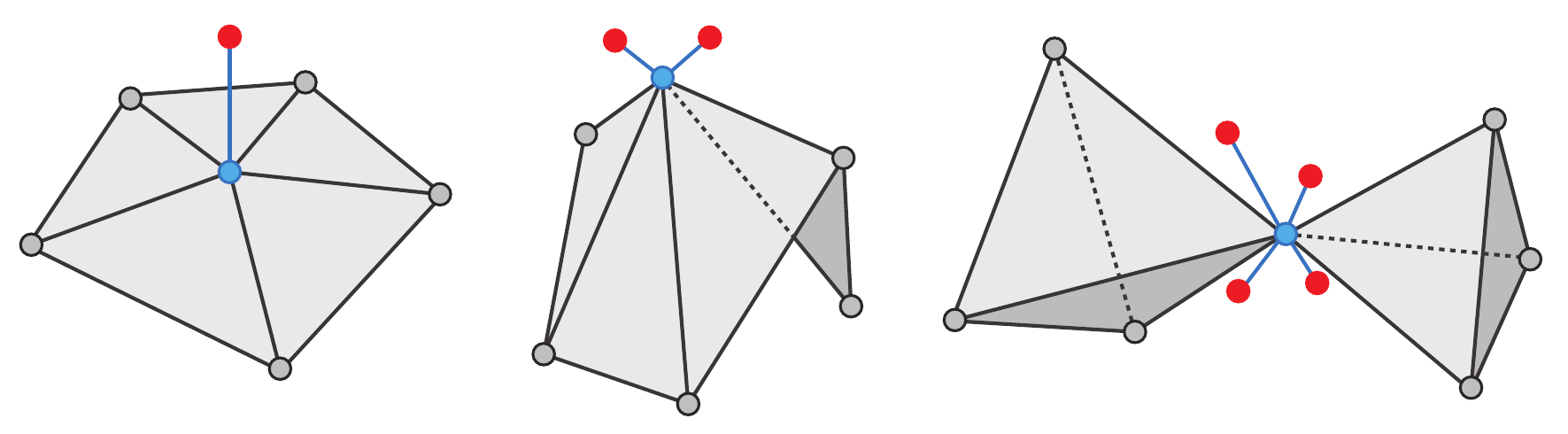}
\vspace{-2em}
\caption*{\label{fig:vertexoffset} {A vertex (blue) of $\MI$ can have one or more offset points (red).}}
\vspace{-1.0em}
\end{wrapfigure}describe how to tackle the general scenario, i.e. multiple offset points, via dynamic programming. 

\paragraph{linear constrained quadratic optimization} Let $O_i$ be $V_i$'s unique offset, we assume the local region of $V_i$ is formed by a triangle set $\mathcal{T}_i$ where every triangle $T_j \in \mathcal{T}_i$ is equipped with an offset distance $d_j$. 
%We then take the average of all $d_j$ values, denoted as $d$, to represent the offset distance for $V_i$. 
Denote $N = |\mathcal{T}_i|$. Note that, since we assume the input could have an arbitrary topology, we set $\mathcal{T}_i$ to be the set of triangles that $V_i$ directly adjacent to if the boundary of this triangle set contains a simple circle topology. Otherwise, we consider $\mathcal{T}_i$ as the set of all triangles with an $\epsilon$ distance from $V_i$. We set $\epsilon = 10^{-5}l$ by default, where $l$ is the length of the diagonal of $\MI$'s bounding box.

By requiring the point-to-plane distance from $O_i$ to $T_j \in \mathcal{T}_i$ to be $d_j$, we can compute $O_i$ by solving the following optimization:
\begin{equation} \label{eq:quadratic}
\underset{O_i}{\text{argmin}}\; ||O_i - V_i||^2 \quad \text{s.t.}\; \text{P}_j^\text{T}O_i = d_j \quad\forall j<N,
\end{equation}
where $\text{P}_j =[a\text{ }b\text{ }c\text{ }d]^\text{T}$ represents the plane of triangle $T_j$ defined by the equation $ax + by + cz + d = 0$ and $a^2 + b^2 + c^2 =1$. The optimization term $||O_i - V_i||^2$ is introduced to pick the position of $O_i$ nearest to $V_i$ when the solution space is a plane or a line. While \prettyref{eq:quadratic} may be solved using OSQP solver \cite{osqp}, due to numerical precision issues, we solve it as a two step process. We first solve an unconstrained quadratic problem:
\begin{equation} \label{eq:quadratic}
\underset{O_i}{\text{argmin}}\; \lambda||O_i - V_i||^2 + \sum_{j}{(\text{P}_j{^\text{T}}O_i - d_j)^2} \quad\forall j<N.
\end{equation}
The role of \(\lambda\) is to ensure that, in cases where there are infinitely many solutions satisfying the point-to-plane distances, an optimal solution can be selected, while minimizing its impact on other scenarios. Consequently, it is necessary to set it to a very small number. By default, we set \(\lambda\) to $10^{-9}$. Subsequently, this expression can be solved using the least squares method. We employ the QR decomposition from the Eigen library for the solution.

We then perform a distance check for each plane to ensure $|\text{P}_j^\text{T}O_i - d_j| \leq \alpha$, where $\alpha$ is a parameter to control how much the user desired point to plane distance is satisfied. Due to the presence of $\lambda$, the distances between the solved point position and the planes will always be slightly off from the desired values. Therefore, \(\alpha\) cannot be set to zero but must be assigned a small value. We set $\alpha = 10^{-6}$ by default. Since there could be cases that cannot satisfy the distance check for all triangles in $\mathcal{T}_i$, we then propose to address this issue by generating multiple offsets for a vertex as detailed in the next paragraph.

\paragraph{Dynamic Programming}
When there is no solution found for \prettyref{eq:dynamic_update}, we compute multiple offset points. We propose to solve this issue by dividing the triangle set $
\mathcal{T}_i$ into $K$ subsets, denoted by $\mathcal{Q} = \{\mathcal{T}_i^0, \mathcal{T}_i^1, \cdots\}$ where $|\mathcal{Q}| = K$, and we compute an offset point $O_i^k$ for each subset $\mathcal{T}_i^k$. Denote $\mathcal{R}$ as the offset point set, i.e. $\{O_i^0, O_i^1, \cdots\}$ and $|\mathcal{R}| = K$. Therefore, we need to solve the following problem:
\begin{equation}
    \begin{split}
    \label{eq:dynamic_update}
\underset{\mathcal{Q}, \mathcal{R}}{\text{argmin}}\;&  \lambda \sum_{k} ||O_i^k - V_i||^2 + \sum_{k}{(\text{P}_j^k{^\text{T}}O_i^k - d_j^k)^2}\\
    \end{split}
\end{equation}

The challenge arises for deciding the $K$ subsets while ensuring the point-to-plane distance constraint for the planes of $\mathcal{T}_i^k$ for each $O_i^k$. While we can simply decompose $
\mathcal{T}_i$ based on the similarity of the normal of the triangles, solution of $O_i^k$ is still not guaranteed to exist and normal could be wrongly computed when floating point precision is involved. Instead, we propose a dynamic programming strategy to solve $\mathcal{Q}, \mathcal{R}$ as detailed below, where \prettyref{alg:dp} is the pseudo-code.

Given the $N$ triangles in $\mathcal{T}_i$, a subset could have $2^N$ possible configurations. We encode each subset using the binary representation with length $N$. For example, $0b010101$ represents the subset of $\{T_0, T_2, T_4\}$ for $N=6$, while $0b100100$ is the subset of $\{T_2, T_5\}$. We use $D_o$ to store the energy values of \prettyref{eq:quadratic} when solving an offset vertex $O_i^k$ for the group of triangles $\mathcal{T}_i^k$ represented in the aforementioned binary format $x$, and $D_c$ to store the coordinates of the corresponding offset vertex $O_i^k$. As shown in \prettyref{alg:dp}, we employ a recursive scheme (lines 2-12) to find the best group decomposition, where for each level of the recursion we decompose a set $x$ into two subsets $\Tilde{x}$ and $x - \Tilde{x}$ if $x$ cannot be solved by \prettyref{eq:quadratic}. Here, the union of the triangles represented by $\Tilde{x}$ and $x - \Tilde{x}$ is the same as the triangle set of $x$. 
In this case, we use the an array of scalars, $D_s$, to record the best decomposition for $x$, i.e. $D_s[x] = \Tilde{x}$. From which, we can retrieve all the decomposed subsets of $\mathcal{T}_i$, as shown by lines 15-20 of \prettyref{alg:dp}. Note that, the dimension of $D_o$, $D_c$, and $D_s$ are all $2^N-1$. 
\begin{wrapfigure}{r}{0.5\linewidth}
\vspace{-0.5em}
\flushright
\includegraphics[width=\linewidth]{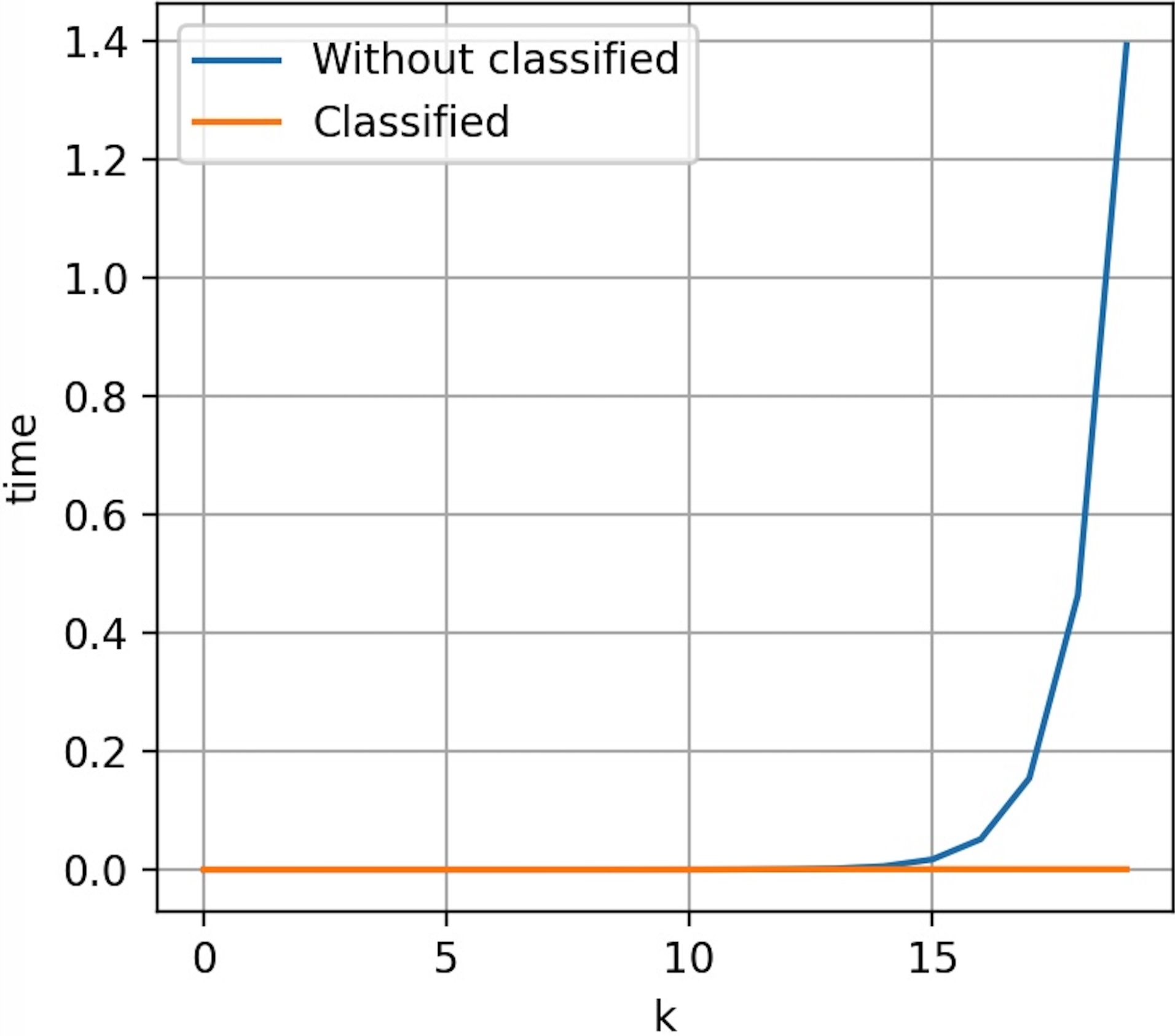}
\vspace{-2em}
\caption{\label{fig:speedforclassify} {The two curves respectively demonstrate the computational time required to calculate a point with and without prior classification. It can be observed that when \( K \) is particularly large, lacking this classification makes it difficult to compute the offset points within a reasonable timeframe.}}
\vspace{-1.0em}
\end{wrapfigure}
The introduced dynamic programming has a computational complexity of $O(3^k+2(1-2^k))$ \cite{giraudo2015combinatorial,A101052}. The computation is efficient for $N\leq 12$, e.g. $\leq 0.01s$ without any parallelization. Since the dynamic programming  operates independently in each local context, it can be fully parallelized. However, if $N$ becomes excessively large, the computation time may grow significantly as show in \prettyref{fig:speedforclassify}.

To address this problem, we first classify the the planes defined by $\mathcal{T}_i$ into groups where the planes within a group have the similar normal. We then compute an averaged plane for each group and consider the averaged one as the plane for all the triangles with each group.

\begin{figure}[t!]
\centering
\includegraphics[width=\linewidth]{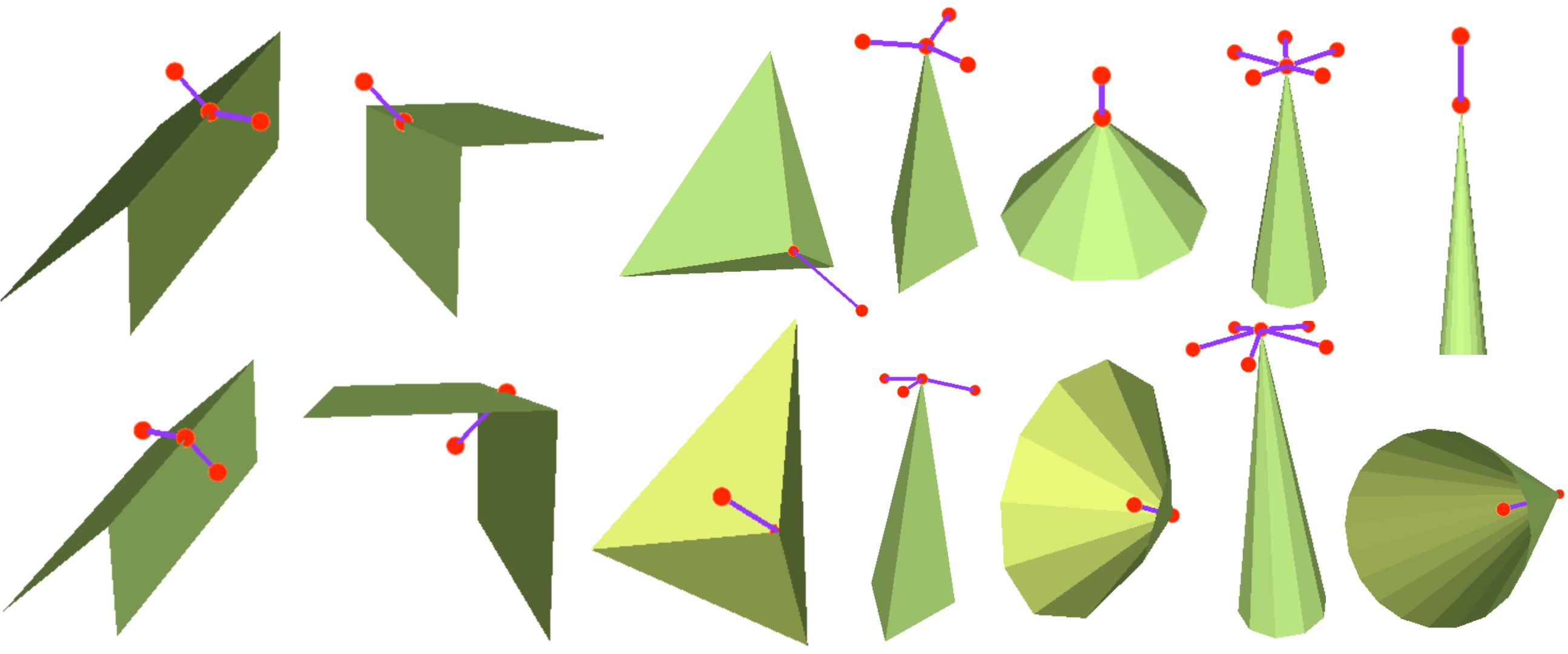}
\caption{\label{fig:multi-offsets} Nine different vertex offset scenarios computed by our approach on a mesh example. Each column corresponds to one scenario, where the top and bottom rows show two different offset direct (outward and inward). It can be observed that in some sharp cases, the points generated by the inward offset of vertices may appear outside the input model. This situation does not pose any issues. The subsequent convex hull and winding number computations will exclude these points and .}
\end{figure}

%Additionally, this strategy, when applied to regions that lack sufficient smoothness, can lead to an improvement in the quality of the generated offset mesh. This effect can be observed in Figure \ref{fig:merge_diff}. In other words, the process is bifurcated into two distinct phases.
To summarize, we compute the offsets for each vertex of $\MI$ by firstly merging its adjacent triangles with similar normals, and then perform \prettyref{alg:dp} to solve the offset points. \prettyref{fig:multi-offsets} illustrates different offset cases on a real example.

% \begin{figure}[ht!]
% \centering
% \includegraphics[width=\linewidth]{fig/merge_rad.pdf}
% \caption{\label{fig:merge_diff}When the angle between the normal vectors of two facets is less than \(\theta\), these two facets will be merged. Taking an offset distance of $1\%$ of the diagonal length directed inward as an example, the first row represents the original model, \(\theta = 1\), \(\theta = 10\). The second row shows \(\theta = 30\), \(\theta = 50\), \(\theta = 100\), respectively.}
% \end{figure}

\SetNlSty{textbf}{}{:}
\begin{algorithm}
\SetAlgoLined
\DontPrintSemicolon
\caption{\label{alg:dp}Dynamic Programming for Vertex Offset}
    \KwIn{$V_i$, $\mathcal{T}_i$, $d$, $T_j\in\mathcal{T}_i$}    \KwOut{$\mathcal{Q}$, $\mathcal{R}$\;
}
Set each entry of $D_{o}$, $D_{c}$, and $D_{s}$ as \text{NULL}\;
\SetKwProg{Fn}{Function}{:}{end}
\Fn{DFS(x)}{
    \If{$D_{o}[x] \neq \text{NULL}$}{
             \Return{$D_{o}[x]$}
        }
    $D_{o}[x] \gets \infty$\;
    \If{Eq.\ref{eq:quadratic}($x$) is solvable}{
             Set Eq.\ref{eq:quadratic}($x$)'s energy and solution to $D_{o}[x]$ and $D_{c}[x]$
        }
        \For{each $\Tilde{x}$ derived from $x$}{
            $o \gets$ \textit{DFS($\Tilde{x}$)} + \textit{DFS($x - \Tilde{x}$)}\;
            \If{$o < D_{o}[x]$}{
                    $D_{o}[x] \gets o$, $D_{s}[x] \gets \Tilde{x}$\;
            }
        }
        \Return {$D_{o}[x]$}
    }
    \textit{DFS($2^N-1$)}\;
    $\Tilde{\mathcal{Q}}\gets\emptyset$, $\mathcal{R}\gets\emptyset$, $\Tilde{\mathcal{Q}}.push(2^N-1)$\;
    \While{$\Tilde{\mathcal{Q}} \neq \emptyset$}{
         $x \gets \Tilde{\mathcal{Q}}.front()$,  $\Tilde{\mathcal{Q}}.pop()$\;
         \eIf{$D_s[x] = \text{NULL}$}{
         $\mathcal{Q}.insert(x)$, $\mathcal{R}.insert(D_c[x])$\;
        }{
         $\Tilde{\mathcal{Q}}.push(D_s[x])$, $\Tilde{\mathcal{Q}}.push(x - D_s[x])$\;
        }
    }
\end{algorithm}

\subsection{Local Offset Polyhedron}
\label{sub:to}
% \begin{figure}[t]
% \centering
% \includegraphics[width=\linewidth]{fig/polyhdron.pdf}
% \caption{\label{fig:polyhdron} {Given a triangle and the offset points of its vertices (left), the offset volume of this triangle can be either formed through direct offsetting (middle) triangle and convex hull (right).}}
% \end{figure}

\begin{figure}[h!]
\centering
\includegraphics[width=\linewidth]{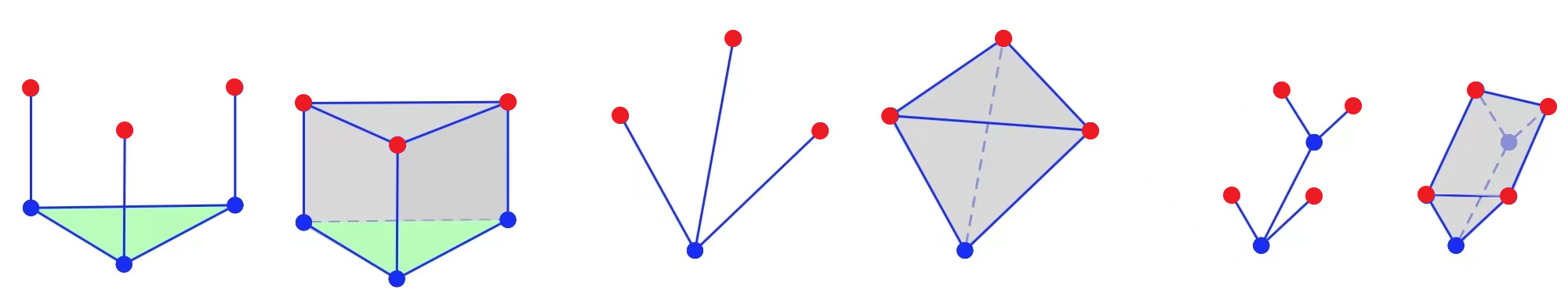}
\caption{\label{fig:polyhedron} {From left to right, the diagrams respectively illustrate the construction of offset volumes from a triangle, a vertex, and an edge.}}
\end{figure}

After generating the offset points of $\MI$, this step is to generate the local offset volumes for the elements of $\MI$, i.e. vertices, edges, and triangles. 
Ideally, if there is only one offset point for each vertex, then we only need to generate the local offset volume $P_i$ for each triangle, $T_i\in\MI$, and $P_i$ is simply the triangle prism formed by $T_i$'s three vertices and their corresponding offset points as illustrated in \prettyref{fig:polyhedron} left. 
% \setlength{\columnsep}{10pt}
% \begin{wrapfigure}{r}{0.7\linewidth}
% \vspace{-0.5em}
% \flushright
% \includegraphics[width=\linewidth]{fig/polygon3}
% \vspace{-2em}
% \caption*{\label{fig:polyhdron} {From left to right, the diagrams respectively illustrate the construction of offset volumes from a triangle, a vertex, and an edge.}}
% \vspace{-1.0em}
% \end{wrapfigure}

In practice, however, one vertex may have multiple offset points, the offset volume computation for $\MI$ would be slightly involved. We tackle the problem by computing an offset polyhedron for each vertex, edge, and triangle of $\MI$ respectively. For each element, its offset polyhedron is computed as the convex hull of a point set associated with the element. To be specific, for a vertex $V_i$ of $\MI$, the point set includes $V_i$ and all its offset points $\{O_i^0, O_i^1, \cdots\}$ corresponding to the neighboring triangle set $\mathcal{T}_i$ (\prettyref{fig:polyhedron} middle). For an edge $E_i = \{V_0, V_1\}$ of $\MI$, its polyhedron is computed as the following (\prettyref{fig:polyhedron} right). Assume $E_i$ is adjacent to a set of triangles $\mathcal{T}_{E_i}$, the point set of $E_i$ for the convex hull computation includes $V_0, V_1$, and those from $V_0$'s and $ V_1$'s offset points satisfying the following condition: the offset point's corresponding triangle subset solved by \prettyref{alg:dp} contains any triangle of $\mathcal{T}_{E_i}$. Similarly, for a triangle $T_i = \{V_0, V_1, V_2\}$ of $\MI$, its polyhedron is the convex hull of the point set including $V_0, V_1, V_2$, and those offset points of $V_0, V_1, V_2$ as long as the offset point's corresponding triangle subset solved by \prettyref{alg:dp} contains $T_i$.

This strategy leads to a set of polyhedra, denoted as $\mathcal{P}$, that their union forms the offset volume of $\MI$.

\subsection{Geometry Extraction}
\label{sub:ge}
% \begin{figure}[t]
% \centering
% \includegraphics[width=\linewidth]{fig/fgroups.pdf}
% \caption{\label{fig:fgroups} A face of $\MC$ is with one of the illustrated five types. Left: $\text{T}_\text{II}$ (blue), $\text{T}_\text{IV}$ (gray), $\text{T}_\text{V}$(yellow); middle: $\text{T}_\text{I}$ (green); right: $\text{T}_\text{III}$ (red).}
% \end{figure}

After generating the offset polyhedra, this section is to compute all the triangles that constitute the geometry of $\MO$ in three steps. 
\setlength{\columnsep}{10pt}
\begin{wrapfigure}{r}{0.4\linewidth}
\flushright
\vspace{-1.5em}
\includegraphics[width=\linewidth]{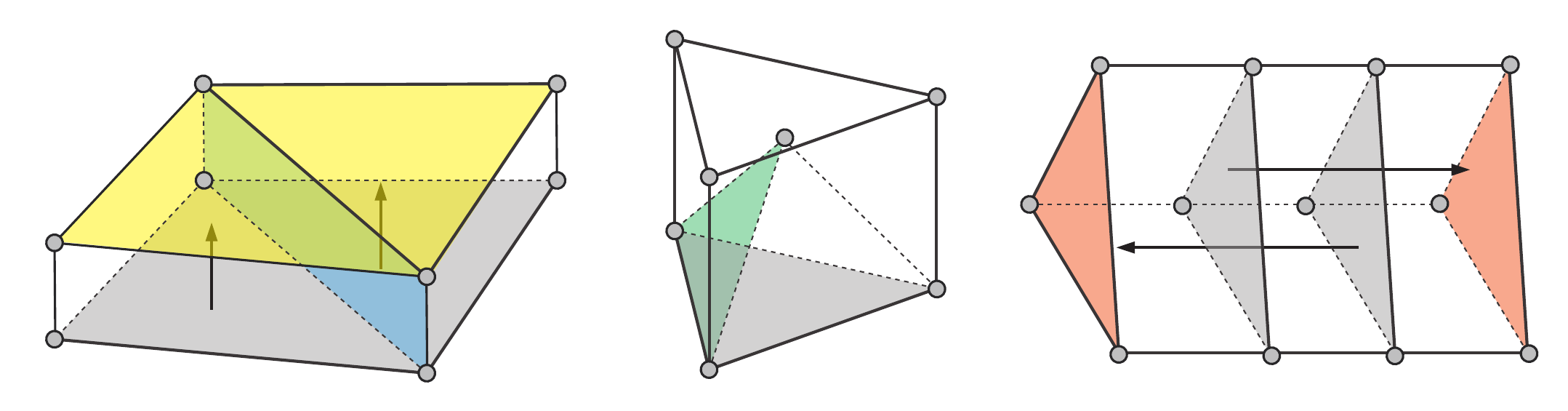}
\vspace{-2em}
\caption*{\label{fig:fgroups} A face of $\MC$ is with one of the illustrated five types. Left: $\text{T}_\text{II}$ (blue), $\text{T}_\text{IV}$ (gray), $\text{T}_\text{V}$(yellow); middle: $\text{T}_\text{I}$ (green); right: $\text{T}_\text{III}$ (red).}
\vspace{-1.5em}
\end{wrapfigure}
% The main workflow of Geometry Extraction consists of three steps, and prior to executing these primary steps, we implement a pre-classification strategy to accelerate the subsequent steps. We divide all the triangles from the union of all the polyhedra into two categories, \(M_X\) and \(M_Y\). If a triangle has at least one vertex on \(M_I\), then it will be classified into \(M_X\); otherwise, it will be classified into \(M_Y\). Triangles in \(M_X\) are more likely to become components of \(M_O\), while those in \(M_Y\) have a much lower probability of becoming components of \(M_O\).
First, we convert the all triangles from $\mathcal{P}$ into a conforming triangle mesh \(M_C\) by resolving all the intersections, where the intersection points and line segments are taken as constraints for a subsequent Delaunay triangulation for each face of $P_i \in \mathcal{P}$. After this step, \(M_C\) is intersection-free and contains only triangles. As illustrated in the inset, a triangle of \(M_C\) can be classified into one of the five types, i.e. enclosed by a $P_i$ ($\text{T}_\text{I}$), shared by two $P_i$s that at different sides of the triangle ($\text{T}_\text{II}$), has an opposite sign to the offset direction as evaluated by $\MI$ ($\text{T}_\text{III}$), a triangle of $\MI$ ($\text{T}_\text{IV}$), and a face of $\MO$ ($\text{T}_\text{V}$). Since $\text{T}_\text{IV}$ faces are properly tracked from the beginning, our next two steps are to obtain the triangles with type $\text{T}_\text{V}$ by filtering out those triangles belonging to the first three types.

Second, we identify and discard triangles with $\text{T}_\text{I}$ and $\text{T}_\text{II}$ through rational number represented ray intersection checks where the rays are shot from the centers of these faces along their normal directions. Both types can be easily detected by checking if there are any intersections between the rays and other polyhedra other than the ones these faces belong to. 

Third, we detect faces with $\text{T}_\text{III}$ through the generalized winding number \cite{jacobson2013robust}. Since the winding number computation is not exact, when the offset distances $d$ are close to the numerical precision, $\text{T}_\text{III}$ faces may be wrongly labeled, resulting redundant faces in $\MO$. However, accordingly to our extensive experiments, we don't observe  such an issue for $d\geq 10^{-6}$ for both uniform and varying offsets. Note that, if the input mesh is watertight and free of self-intersections, $\text{T}_\text{III}$ faces can be filtered out completely by employing \cite{cgal:eb-23b}.
% Once we obtain \(\MC\), i.e., the triangle soup composing the geometry of \(\MO\). We can observe that there are some holes on \(\MC\), 
% which are due to the fact that some facets from \(M_Y\) may also become part of \(\MO\), but we did not compute them in the previous step due to their low probability.
% At this point, we need to first detect all the edges of the holes and then search for the faces on \(M_Y\) that coincide with these edges. We perform Geometry Extraction again on these faces. In very rare cases, we find that there are still undesired holes present, so we proceed to search for boundaries, query \(MY\), and execute Geometry Extraction once more.

\subsection{Topology Construction}
\label{sub:tc}
After obtaining \(\MC\), i.e., the triangle soup composing the geometry of \(\MO\), in this section, we accomplish the connectivity of $\MO$ to make it free of degenerates and intersections under exact number representations. $\MO$ is also watertight given that the input is watertight and self-intersection-free. 

We execute three steps to convert \(\MC\) to \(\MO\). First, we merge the vertices with the same coordinates in rational numbers so that all triangles are connected but zero-area holes may still exist. Second, we perform a hole-filling to remove the zero-area gaps. Given a boundary loop, our hole-filling is done by inserting an edge at a time with two criteria: this edge has no intersection with any other triangles and the areas of the newly generated polygons are still zero, until the loop is fully triangulated. Third, we eliminate all degenerate elements generated during the hole-filling step through edge collapse and edge split without changing the geometry of \(\MO\).

\subsection{Speedup Strategies}
We employ speedup strategies from different aspects to improve the performance of our approach as detailed below. 

We use OpenMP to trivially parallelize the vertex offset generation (\prettyref{sub:vo}) and the triangle polyhedra construction (\prettyref{sub:to}). Furthermore, we spatially partition the spatial domain, which is covered by the geometry formed by extending \(\MI\) along three axes with the largest offset distance, into cubes. Each cube is associated with the list of triangles of \(\MI\) that are either fully contained or have intersections with the cube. So the computational domain of \(\MO\) is constrained within each cube. We apply this strategy for \prettyref{sub:ge} and \prettyref{sub:tc}.

Data-structure-wise, we employ AABB-tree data-structure as much as possible for spatial  search of vertex, face, and polyhedron elements, such as the $\mathcal{T}_i$ collection for $V_i$ in \prettyref{sub:vo}, the polyhedron neighborhood search during intersection resolving, the ray-intersection check and the winding number computation during invalid triangle filtering in \prettyref{sub:ge}. 
\begin{wrapfigure}{r}{0.4\linewidth}
\flushright
\vspace{-1.0em}
\includegraphics[width=\linewidth]{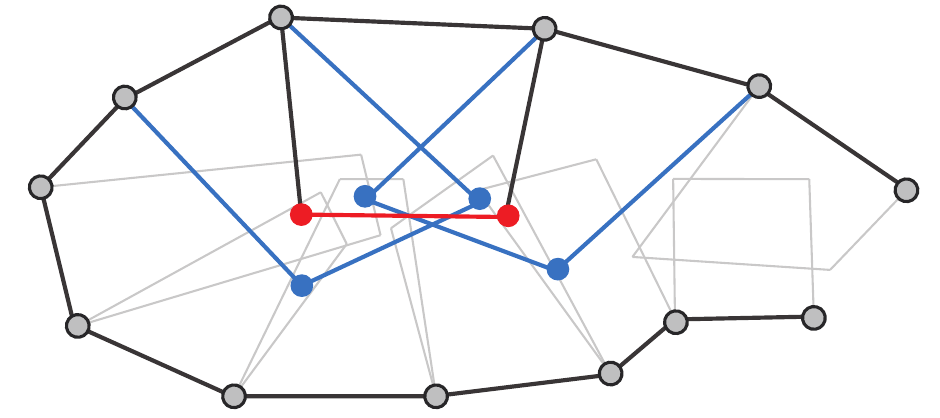}
\vspace{-1.0em}
\caption*{While many polygons intersect with the edge in red, two polygons in blue are enough to early reject the edge from being further decomposed into $\text{T}_\text{II}$ edges.}
\vspace{-1.0em}
\end{wrapfigure}
Based on the performance statistics, we find that the most time-consuming part is during the intersection resolving in \prettyref{sub:ge} where an extensive number of exact computations \cite{boolean2007} are involved. While all intersections among the faces of the polyhedra of $\mathcal{P}$ are resolved by splitting the faces into intersection-free triangles, many of the resulting triangles are not part of \(M_O\). In another words, if the entire or the most region of a face will be discarded for computing \(M_O\), then most of the computations for resolving the intersections with this face are wasted since most of the resulting triangles will be discarded as well. Based on the principal that postponing unnecessarily intersection computations as much as possible, we propose a speedup strategy of \prettyref{sub:ge} with the pseudo-code given by \prettyref{alg:multi_speed_up}.

Assume $\mathcal{T}$ is the set of triangles of all polyhedra of $\mathcal{P}$ and $\mathcal{T}_o$ is the set of all triangles of \(M_O\). We first subdivide $\mathcal{T}$ into two sets $\mathcal{T}_{cur}$ and $\mathcal{T}_{later}$ where they contains triangles with high and low chances to be part of $\MO$ respectively. As shown in lines 2-10 of \prettyref{alg:multi_speed_up}, as an initialization, $\mathcal{T}_{later}$ contains triangles with two scenarios that are both unlikely to be part of $\MO$: sharing a vertex with $\MI$ and is inside (outside) $\MI$ for $s = 1$ ($s = -1$). We then obtain $\mathcal{T}_o$ in an iterative way by constructing the most of it using $\mathcal{T}_{cur}$ (lines 12-23) and gradually refining it using $\mathcal{T}_{later}$ (lines 11 and 24-29). By doing so, the output of \prettyref{alg:multi_speed_up} is ensured to be same as the brute-force approach in \prettyref{sub:ge}. During the construction of $\mathcal{T}_o$ using $\mathcal{T}_{cur}$, 
\setlength{\columnsep}{10pt}

we identify a scenario that for a triangle of $\mathcal{T}_{cur}$ that intersects with many polyhedra, a large part of it may be covered by one or several of those polyhedra. For such faces, there would be many intersection computations to produce $\text{T}_\text{I}$ and $\text{T}_\text{II}$ sub-triangles that are eventually discarded through ray-casting test. The inset illustrates a simplified scenario in 2D. Accordingly, in lines 17-22 of \prettyref{alg:multi_speed_up}, we propose to subdivide $T_i$ into triangles by a subset of all its intersecting polyhedra to avoid unnecessarily more detailed intersection computations. Specifically, given a triangle $T_i$ that has intersections with $N_p\ge 5$ polyhedra denoted by the set $\mathcal{P}_i$, we first subdivde $T_i$ into four smaller triangles, i.e. $T^j_i$ with $j = \{0, 1, 2, 3\}$, using the mid-point subdivision. For each $T^j_i$, we then find the polyhedron from $\mathcal{P}_i$ that contains $T^j_i$ the most by creating several sampling points within $T^j_i$ and checking the number of sampling points contained by the polyhedron. After that, corresponding to $T^j_i$, there will be four polyhedra identified, $P^j_i$ with $j = \{0, 1, 2, 3\}$. We then perform the intersection resolving procedure in \prettyref{sub:ge} for $T_i$ and these four polyhedra to obtain a set of subdivided triangle set of $\Tilde{\mathcal{T}_i}$. After filtering out those triangles in $\Tilde{\mathcal{T}_i}$ that are fully contained in any $P^j_i$, $\mathcal{T}_{cur}$ will be updated.

By experimenting on several models randomly chosen from our testing dataset, we find that the early filtering strategy helps our pipeline achieves $\sim$30\text{x} performance gain in large offset distance e.g. more than $1\%$.

\begin{algorithm}
\SetAlgoLined
\DontPrintSemicolon
\caption{\label{alg:multi_speed_up}Speed up of \prettyref{sub:ge}}
    \KwIn{$\mathcal{T}$, $\mathcal{P}$\;} 
    \KwOut{$\mathcal{T}_o$\;}
Set $\mathcal{T}_{cur}$, $\mathcal{T}_{later}$, and$\mathcal{T}_o$ as $\emptyset$\;
\For{each $T_i \in \mathcal{T}$}{ 
    \If{$s = 1$ and Winding\_number(centroid($T_i$)) $> 0.5$}{
        $\mathcal{T}_{later}$.push($T_i$)\;
    }
    \ElseIf{$s = -1$ and Winding\_number(centroid($T_i$)) $\leq 0.5$}{
       $\mathcal{T}_{later}$.push($T_i$)\;
    }
    \ElseIf{($T_i$) have a vertex on $M_i$}{
       $\mathcal{T}_{later}$.push($T_i$)\;
    }
    \Else{
        $\mathcal{T}_{cur}$.push($T_i$)\;
    }
}
\While{$\mathcal{T}_{cur}\neq\emptyset$}{
    $\mathcal{T}'_{cur} \gets\emptyset$\;
    \For{$T_i \in \mathcal{T}_{cur}$}{
        \eIf{$N_p<5$}{
            $\mathcal{T}'_{cur}$.push($T_i$)\;
        }{
           Subdivide $T_i$ into $T^0_i, T^1_i, T^2_i, T^3_i$\;
           Compute $P^j_i$ containing $T^j_i$ the most, $j = \{0, 1, 2, 3\}$\;
           $\Tilde{\mathcal{T}_i}\gets$ resolving intersections between $T_i$ and $P^0_i, P^1_i, P^2_i, P^3_i$\;
            \For{$t_j \in \Tilde{\mathcal{T}_i}$}{
                \If{$t_j$ is not contained in $P^{0}_{i} || P^{1}_{i} || P^{2}_{i} || P^{3}_{i}$}{
                    $\mathcal{T}'_{cur}$.push($t_j$)\;
                }  
            }
        }
    }
    $\mathcal{T}_o \gets$ perform \prettyref{sub:ge} of $\mathcal{T}'_{cur}$ and $\mathcal{P}$\;
    $\mathcal{T}_{cur}\gets\emptyset$\;
    $\mathcal{B}\gets$ Boundary of $\mathcal{T}_o$\;
    \For{$T_i \in \mathcal{T}_{later}$}{ 
        \If{$T_i$ intersects with $\mathcal{B}$}{
            $\mathcal{T}_{cur}$.push($T_i$)\;
            $\mathcal{T}_{later}$.delete($T_i$)\;
        }
    }
}
\end{algorithm}

%% file: 04-experiments.tex
\section{Experiments}
\defcitealias{cgal:eb-23b}{Minkowski sum 2023}
\defcitealias{portaneri2022alpha}{Alpha Wrap 2022}
\begin{figure}[h!]
\centering
\includegraphics[width=\linewidth]{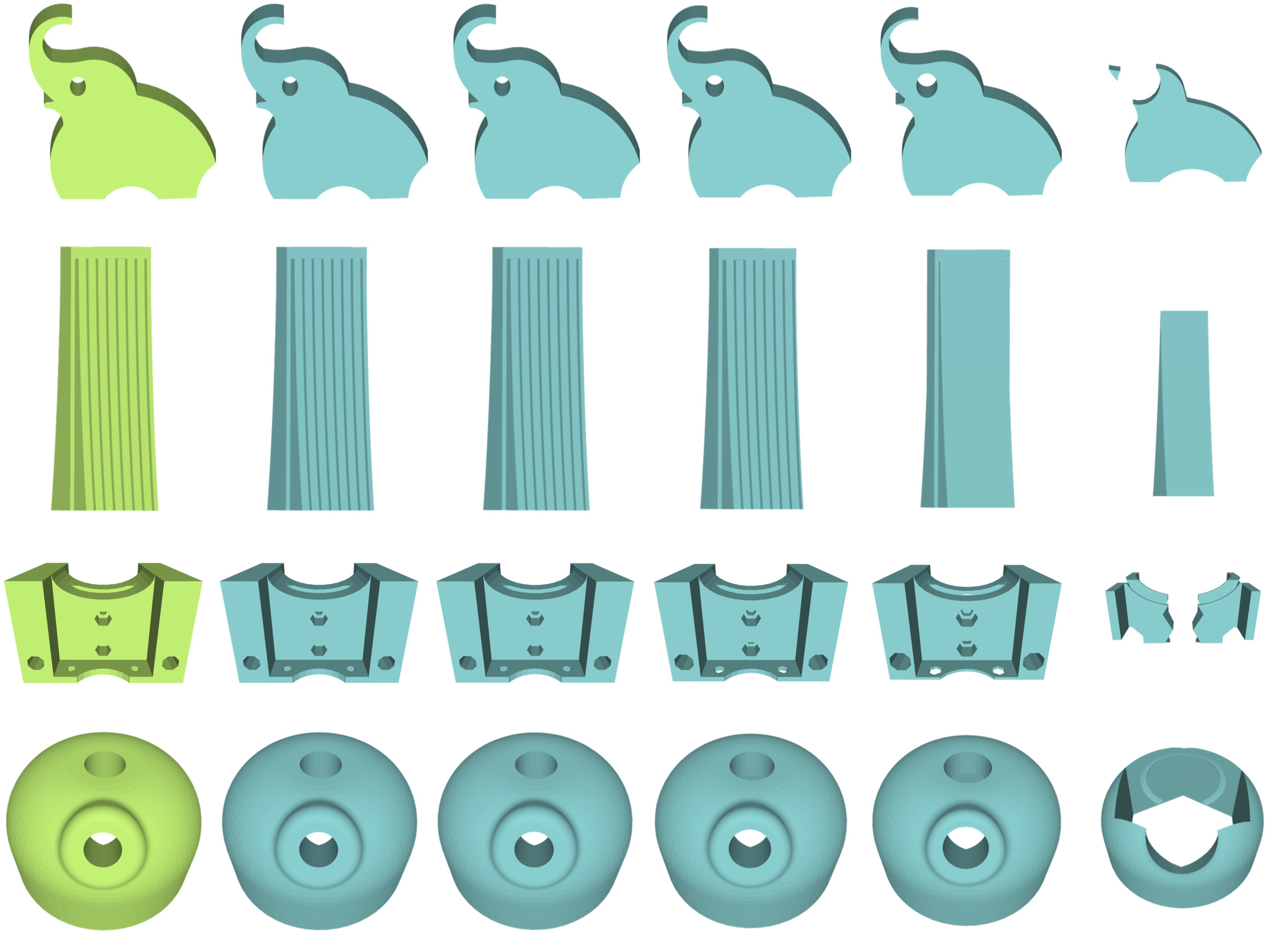}
\caption{\label{fig:showinner}
This figure shows the results of our algorithm for inward offsets at different distances. From left to right, the images are the input, $0.05\%l$, $0.1\%l$, $0.5\%l$, $1\%l$, and $5\%l$.}
\end{figure}

\begin{figure}[h!]
\centering
\includegraphics[width=\linewidth]{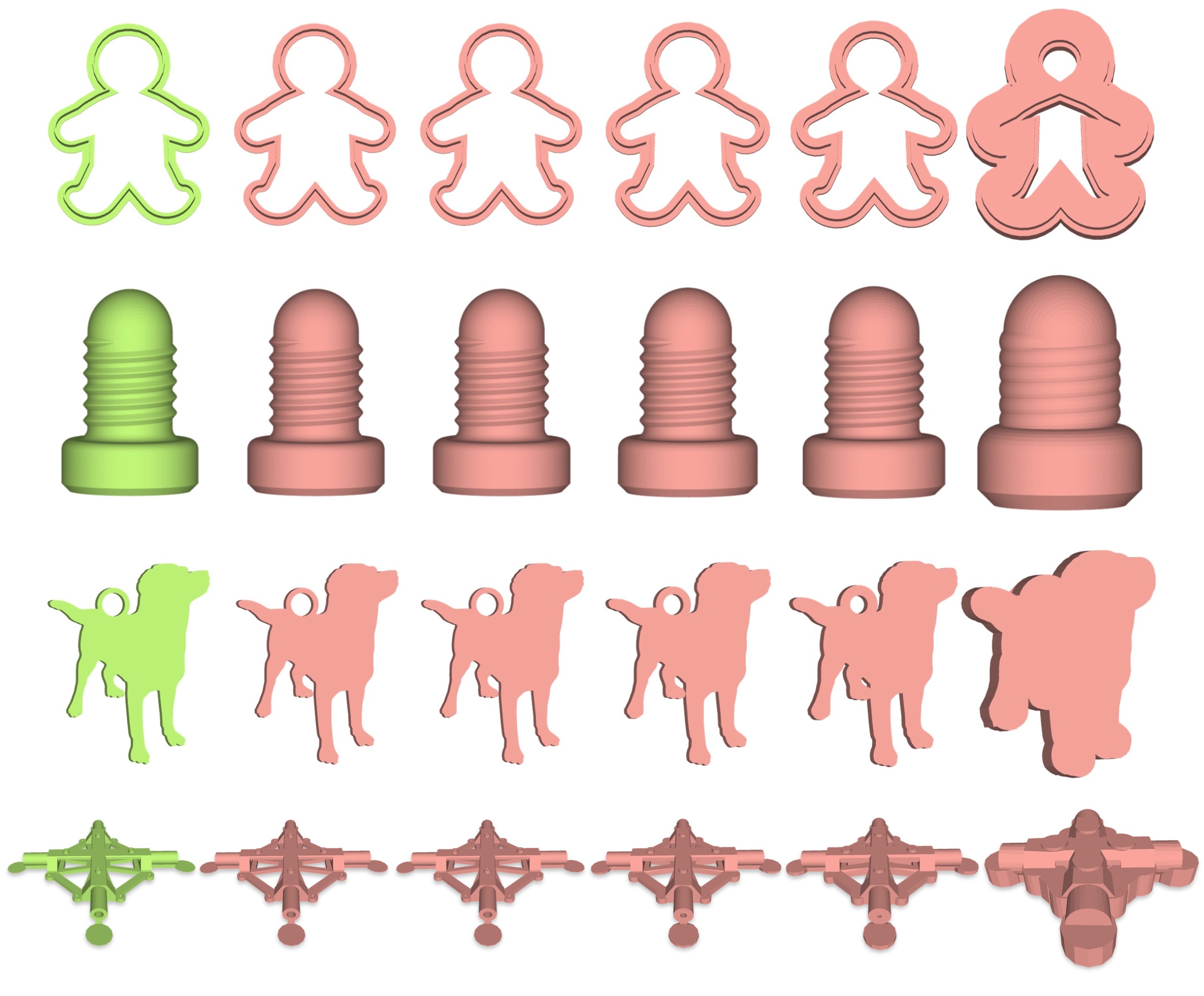}
\caption{\label{fig:showout}
This figure shows the results of our algorithm for outward offsets at different distances. From left to right, the images are the input, $0.05\%l$, $0.1\%l$, $0.5\%l$, $1\%l$, and $5\%l$.
}
\end{figure}
We run all our experiments on a desktop with a 10-cores Intel processor clocked at 5.30 Ghz and 32Gb of memory. We implement our approach in C++, with CGAL \cite{cgal:eb-23b} for exact computations. Throughout the entire paper, our offset distance parameter $d$ is formulated as the ratio of $l$ where $l$ is the length of the diagonal of the input's bounding box, e.g. $d=1\%$ represents that the offset distance is actually $1\%l$. Please refer to the supplementary material for more information.

\subsection{Dataset and Metrics}
\paragraph{Dataset} 
\begin{figure}[t]
\centering
\scalebox{.85}{\includegraphics[width=1.1\linewidth]{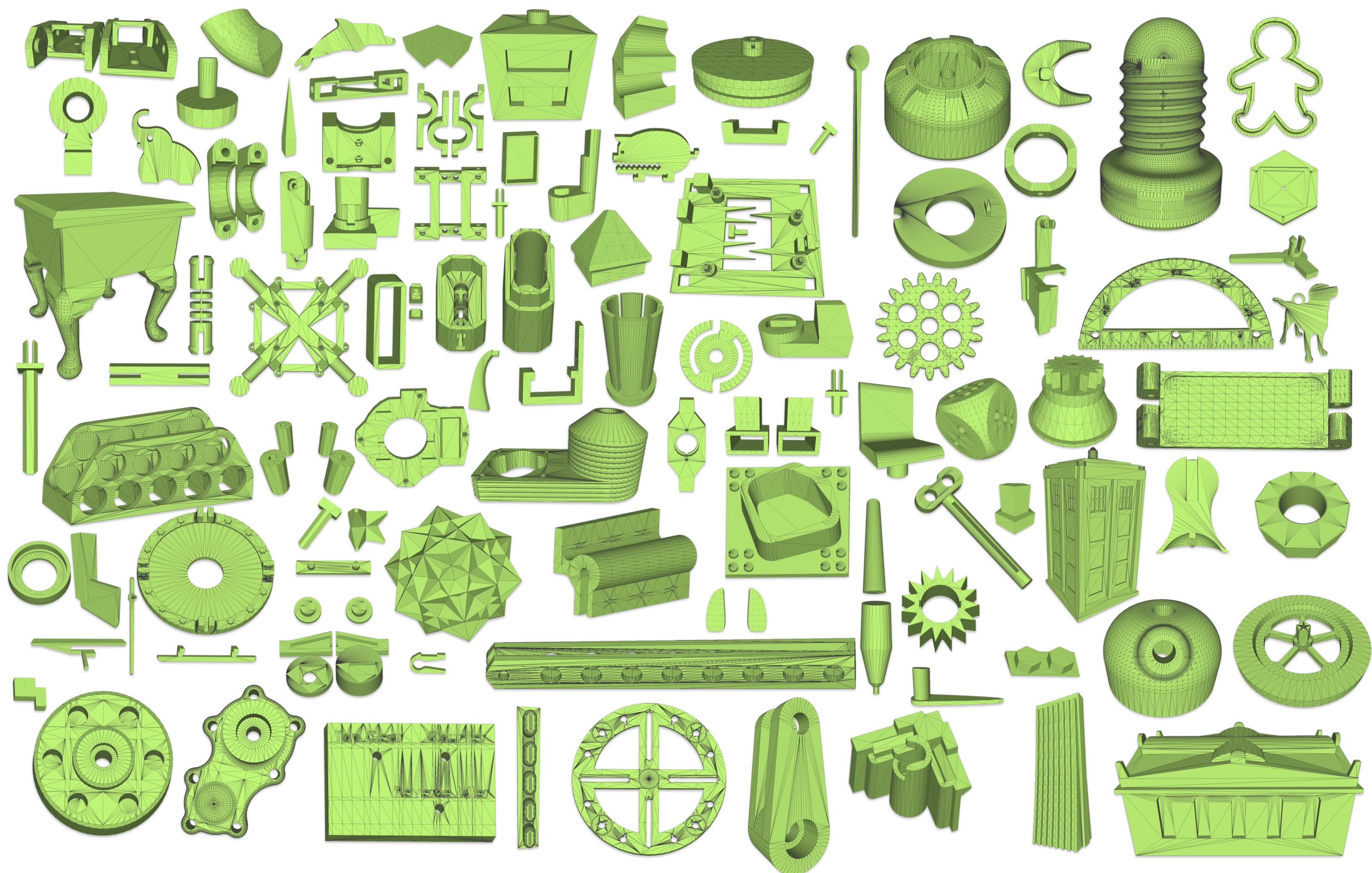}}
\caption{\label{fig:dataset} {A visual gallery of the testing dataset.}}
\vspace{-1.5em}
\end{figure}
We use 100 models chosen from Thingi10K~\cite{Thingi10K} as our test dataset as shown in \prettyref{fig:dataset}. The dataset contains models with varying topology and geometry complexities, In the dataset, there are 50 models with fewer than 1000 triangles. There are 32 models with a triangle count ranging from 1000 to 5000. Additionally, there are 13 models with a triangle count between 5000 and 8000. Lastly, there are 5 models with more than 8000 triangles. 7 models are no manifold. 6 models are with open surface. The average number of genus, disconnected components intersecting triangle pairs and holes are 2.4, 1.94, 13.4 and 0.227, respectively.

% We evaluate the generated offset surfaces from several aspects, including the number of contained triangles, topology (manifoldness) and geometry (self-intersection-free) guarantees, and the geometric distance with the input.

For arbitrary input meshes, our generated offsets are guaranteed to be self-intersection-free and degenerate-free under exact representations. If the input mesh $\MI$ is watertight and self-intersection-free, our generated $\MO$ is guaranteed to be watertight. 

We evaluate the geometry accuracy of our generated offset meshes by measuring how much of the distance between the offset mesh and input is different from the offset distance specified by users. 

\paragraph{Mesh Distances} We employ the typically used point-to-point distance (Hausdorff distance) $H_{point}$ to measure the mesh distance between $\MO$ and $\MI$. Since we aim to generate mitered offset instead of the traditionally constant-radius offset, we also propose to measure point-to-plane distance $H_{plane}$ between $\MO$ and $\MI$. 

Given a vertex $v$ and a triangle mesh $\Tilde{M}$, we denote $d_v(v, \Tilde{M})$ as the closest Euclidean distance from $v$ to $\Tilde{M}$ and $\Tilde{v}\in \Tilde{M}$ is the point on $\Tilde{M}$ closest to $v$. We further denote $d_t(v, \Tilde{M})$ as the Euclidean distance from $v$ to the plane passing through $\Tilde{v}$ and tangent to $\Tilde{M}$. The computations of $H_{point}$ and $H_{plane}$ are summarized in \prettyref{eq:p2p}. 

\begin{equation}
    \begin{split}
 \label{eq:p2p}
H_{point}(M, \Tilde{M})&\triangleq\frac{1}{|M|}\int_{M} d_v(v,\Tilde{M}) ds, \\
H_{plane}(M, \Tilde{M})&\triangleq\frac{1}{|M|}\int_{M} d_t(v,\Tilde{M}) ds,
\end{split}
\end{equation}
$H_{point}(\MO, \MI)$ and $H_{plane}(\MO, \MI)$ measures the average distances from $\MO$ to $\MI$, while $H_{point}(\MI, \MO)$ and $H_{plane}(\MI, \MO)$ are from $\MI$ to $\MO$. The sampling strategy in Metro~\cite{metro} is used for their computation. % We compute the max of the two way distances (\prettyref{eq:h}) when evaluating the geometry distances for the outward offset surface.
% \begin{equation}
%     \begin{split}
%  \label{eq:h}
% H_{point} &= max(H_{point}(\MO, \MI), H_{point}(\MI, \MO)), \\
% H_{plane} &= max(H_{plane}(\MO, \MI), H_{plane}(\MI, \MO)),
% \end{split}
% \end{equation}
% For the inward offset surface, though, we compute the single way distances, i.e. $H_{point} = H_{point}(\MO, \MI)$ and $H_{plane} = H_{plane}(\MO, \MI)$, with the reason being that the inward offset surface may diminish for a large offset distance as shown in \prettyref{fig:varyd} and the computed distances of $H_{point}(\MI, \MO)$ and $H_{plane}(\MI, \MO)$ can be quite off from $d$. 
We use $D_{point} = |H_{point} - d|$ and  $D_{plane} = |H_{plane} - d|$ to indicate how close of our generated mesh aligns with the user specified offset distance, where the smaller of $D_{point}$ and $D_{plane}$ the better. 
\begin{figure*}[h!]
\centering
\includegraphics[width=\linewidth]{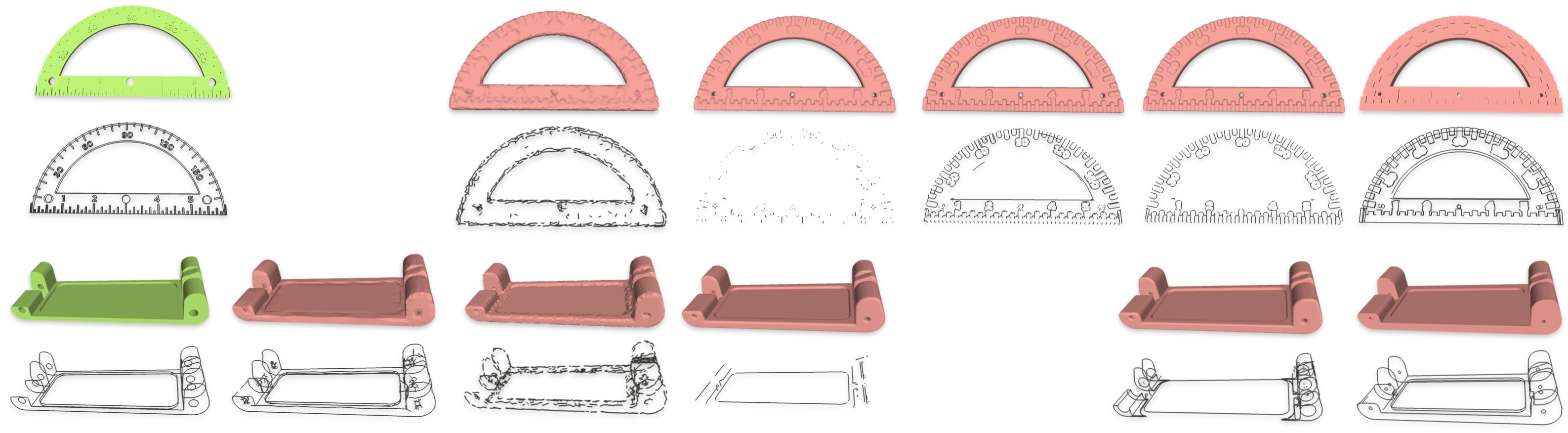}
\footnotesize{
\put(-480,150){Input}
\put(-430,150){\cite{jung2004self}}
\put(-417,110){TIMEOUT}
\put(-350,150){\citepalias{portaneri2022alpha}}
\put(-275,150){\cite{zhen2023}}
\put(-205,150){[\citetalias{cgal:eb-23b}]}
\put(-200,35){UNSUPPORTED}
\put(-127,150){\cite{zint2023feature}}
\put(-45,150){Ours}
}
\caption{\label{fig:feature1}This figure shows an outward offset of $1\%l$. The left side is the input, and the back shows the comparison results of six methods. It can be seen that our method performs better in restoring feature lines.}
\end{figure*}
\begin{figure}[h!]
\centering
\includegraphics[width=\linewidth]{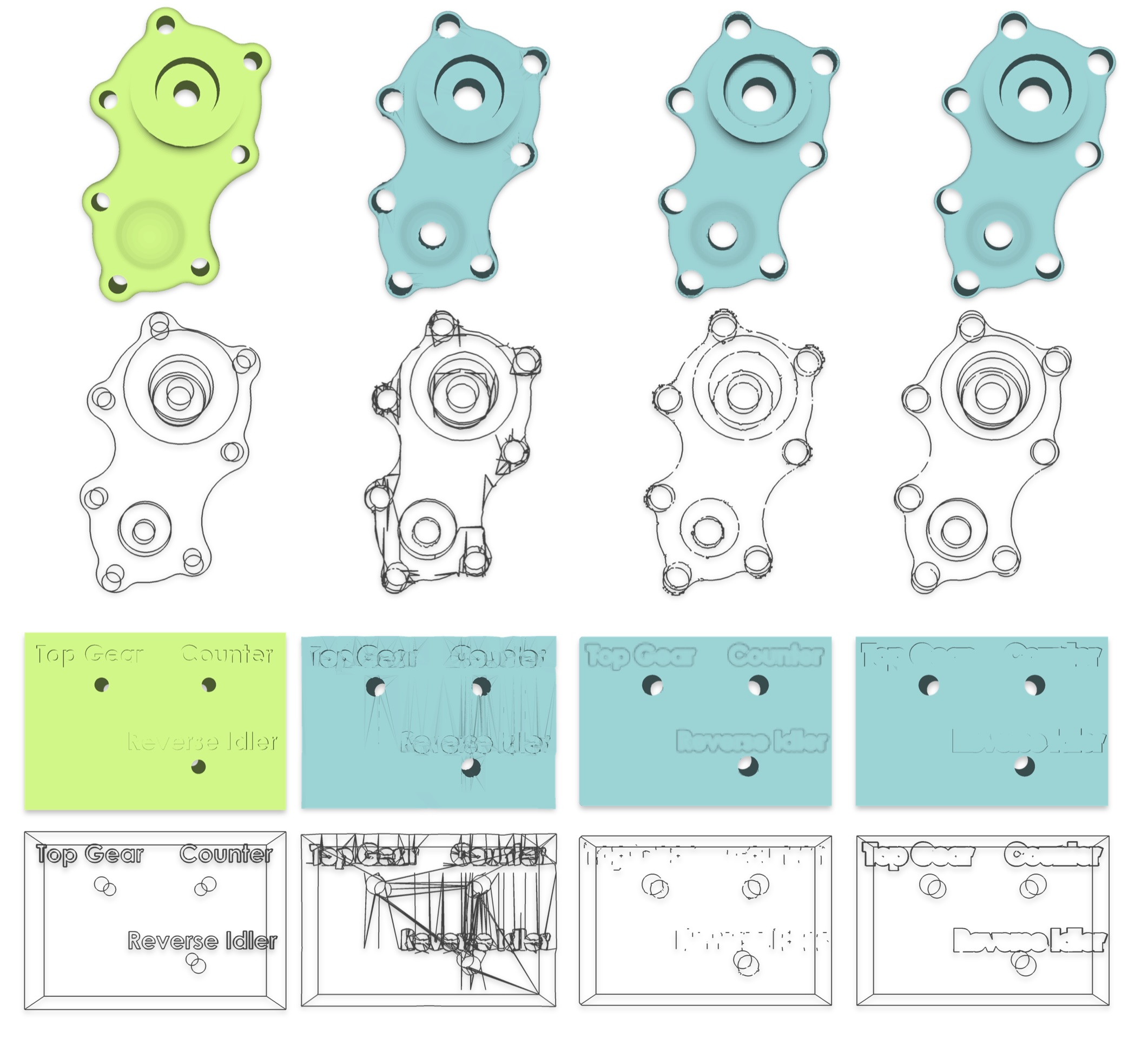}
\footnotesize{
\put(-220,228){Input}
\put(-170,228){\cite{jung2004self}}
\put(-110,228){ \cite{zhen2023}}
\put(-32,228){Ours}
}
\caption{\label{fig:feature_line_in}This figure shows an inward offset of $1\%l$. The left side is the input, and the back shows the comparison results of three methods. It can be seen that our method performs better in restoring feature lines.}
\end{figure}

\paragraph{Feature Preservation} Moreover, to quantitatively measure the effectiveness of various approaches in feature preservation, we propose to compute the difference between sharp features of the input and output meshes. Specifically, based on the simple angle thresholding strategy \cite{gao2019}, we first detect feature lines $S_{\MI}$ and $S_{\MO}$ of $\MI$ and $\MO$ respectively (black lines shown in \prettyref{fig:feature1} and \prettyref{fig:feature_line_in}), then we uniformly sample $S_{\MI}$ and $S_{\MO}$. For each sample $v\in S_{\MI}$, we can find a closest sample $\Tilde{v}\in S_{\MO}$. We denote $d_{angle}(v, S_{\MO})$ as the $l_2$ norm of the dihedral angles of the edges where $v$ and $\Tilde{v}$ belong to.
The difference between $S_{\MI}$ and $S_{\MO}$ can be calculated by 
\begin{equation}
    \begin{split}
 \label{eq:f}
D_{angle}(S_{\MI}, S_{\MO})\triangleq\frac{1}{|S_{\MI}|}\int_{S_{\MI}} d_{angle}(v,S_{\MO}) dl,\\
D_{angle}(S_{\MO}, S_{\MI})\triangleq\frac{1}{|S_{\MO}|}\int_{S_{\MO}} d_{angle}(\Tilde{v},S_{\MI}) dl,
\end{split}
\end{equation}
where $D_{angle}(S_{\MI}, S_{\MO})$ and $D_{angle}(S_{\MO}, S_{\MI})$ measure the average distances from $S_{\MI}$ to $S_{\MO}$ and from $S_{\MO}$ to $S_{\MI}$ respectively.
%compute $D_{angle} = max(D_{angle}(S_{\MI}, S_{\MO}), D_{angle}(S_{\MO}, S_{\MI}))$ for outward offset mesh, and compute $D_{angle} = D_{angle}(S_{\MO}, S_{\MI})$ for inward offset mesh.

\subsection{Our Results} 
\begin{figure}[ht!]
\centering
\includegraphics[width=\linewidth]{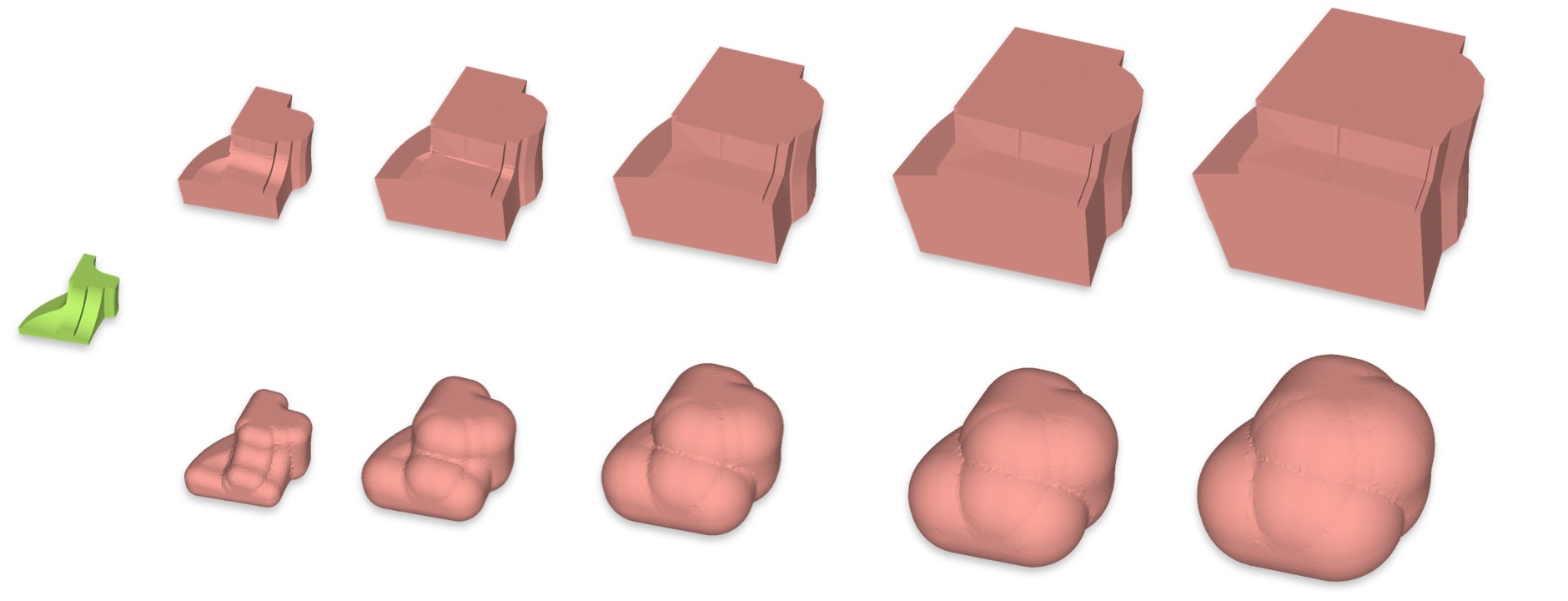}
\caption{\label{fig:baifenzhi}Display the large offset distance. It can be observed that generating an offset with a large offset distance using the original vertices may lead to creases due to the resolution limited by the number of vertices. The offset distance from left to right is $10\%l$, $20\%l$, $30\%l$, $40\%l$, $50\%l$ with outer direct. The first row is our method result, the second row is a case show radius based methods.
}
\end{figure}
Our method requires only a single user-specified parameter, offset distance $d$, and generates the corresponding offset surface mesh that has the following unique properties.

\paragraph{Mitered offset} 

Aiming at generating the mitered offset surface, our approach provides the unique advantage over existing methods by respecting the features of the input excellently, as shown in \prettyref{fig:showinner}, \prettyref{fig:showout}, \prettyref{fig:feature1},
\prettyref{fig:feature_line_in} and
\prettyref{fig:varyd}. 

Unsurprisingly, as demonstrated in \prettyref{fig:baifenzhi}, as the offset distance gets larger, our generated surface tends to be a shape with a rectangular shape, distinct from the constant-radius offset with the tendency of being a sphere.

\begin{figure*}[t]
\centering
{\includegraphics[width=\linewidth]{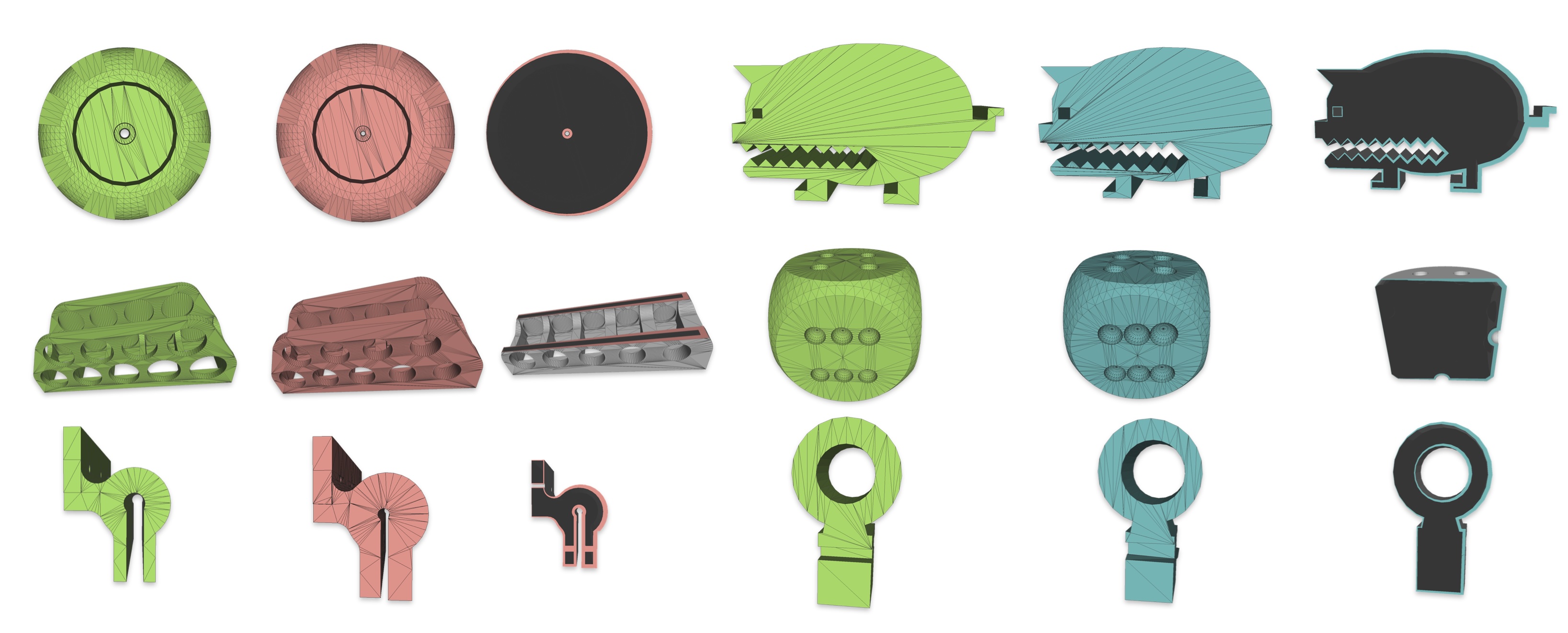}}
\caption{Our approach supports the generation of feature-preserving non-uniform offsets, i.e. from $0.01\%l$ to $2\%l$ for each model, for any given mesh inputs (green), either inwardly (cyan) or outwardly (red).}
\label{fig:non-uniform}
\end{figure*}
\paragraph{Non-uniform offset distance} By varying the offset distance $d$ for different triangles of the input, our approach can easily generate offset meshes with varying offset distances, as illustrated in \prettyref{fig:non-uniform} while maintaining the sharp features of the original meshes. %If the distance around a point changes too drastically, the generated mesh might have some wrinkles. This can be addressed by using the average offset distance of all surrounding faces as the offset distance for each face when calculating the offset distances around each point.

\begin{figure*}[h!]
\flushleft
\setlength{\tabcolsep}{0pt}
\begin{tabular}{cc}
\includegraphics[width=0.99\linewidth]{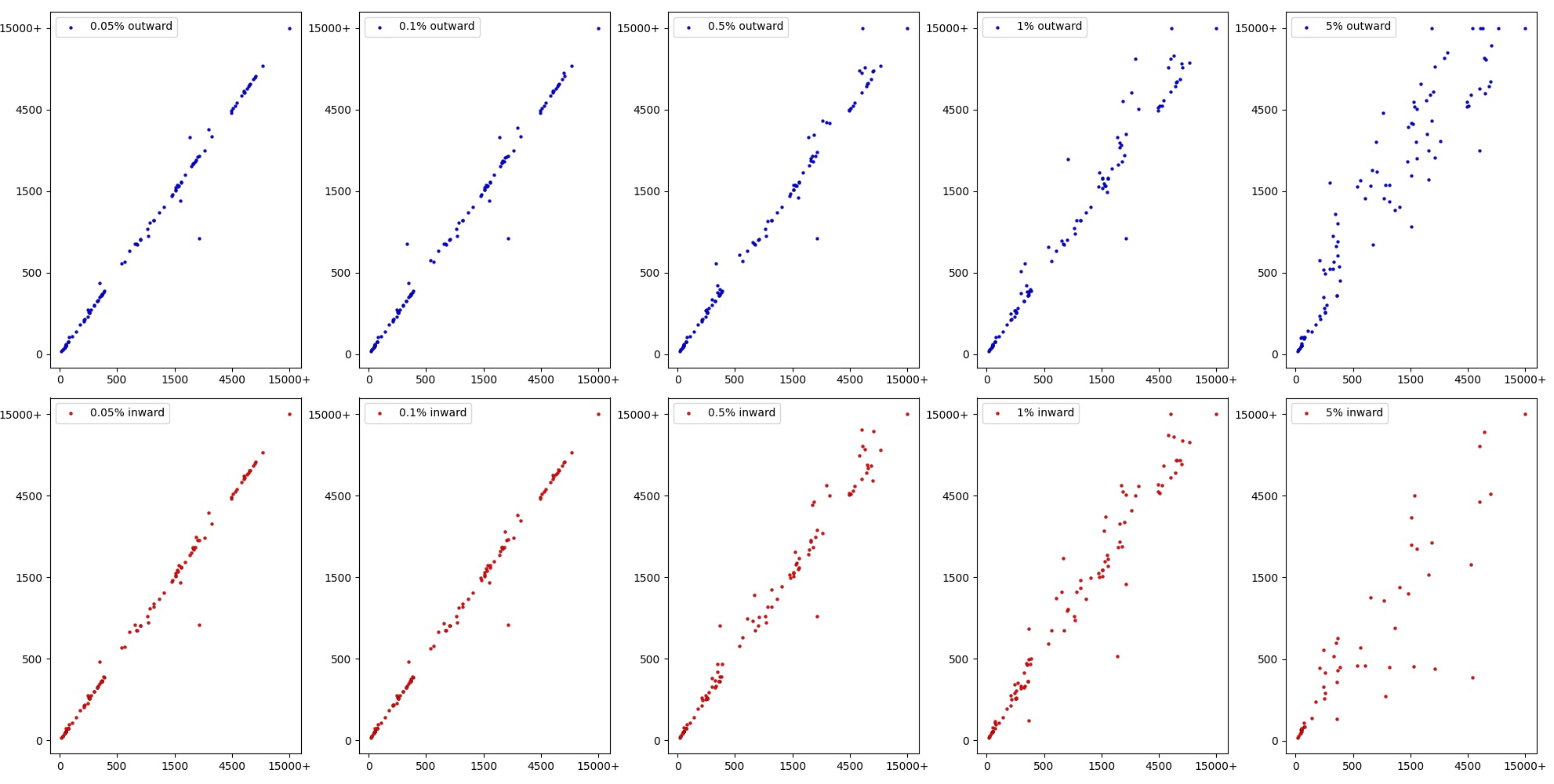}

\end{tabular}
\caption{\label{fig:denity}Triangle number of input VS output generated by our approach.}
\end{figure*}

\paragraph{Topology} 
The topology of our generated mesh is similar to the input mesh, and the density of the mesh is close to or slightly greater than the input mesh as shown in \prettyref{fig:denity}.  

\paragraph{Mesh Repair}
\begin{figure}[t]
\centering
{\includegraphics[width=\linewidth]{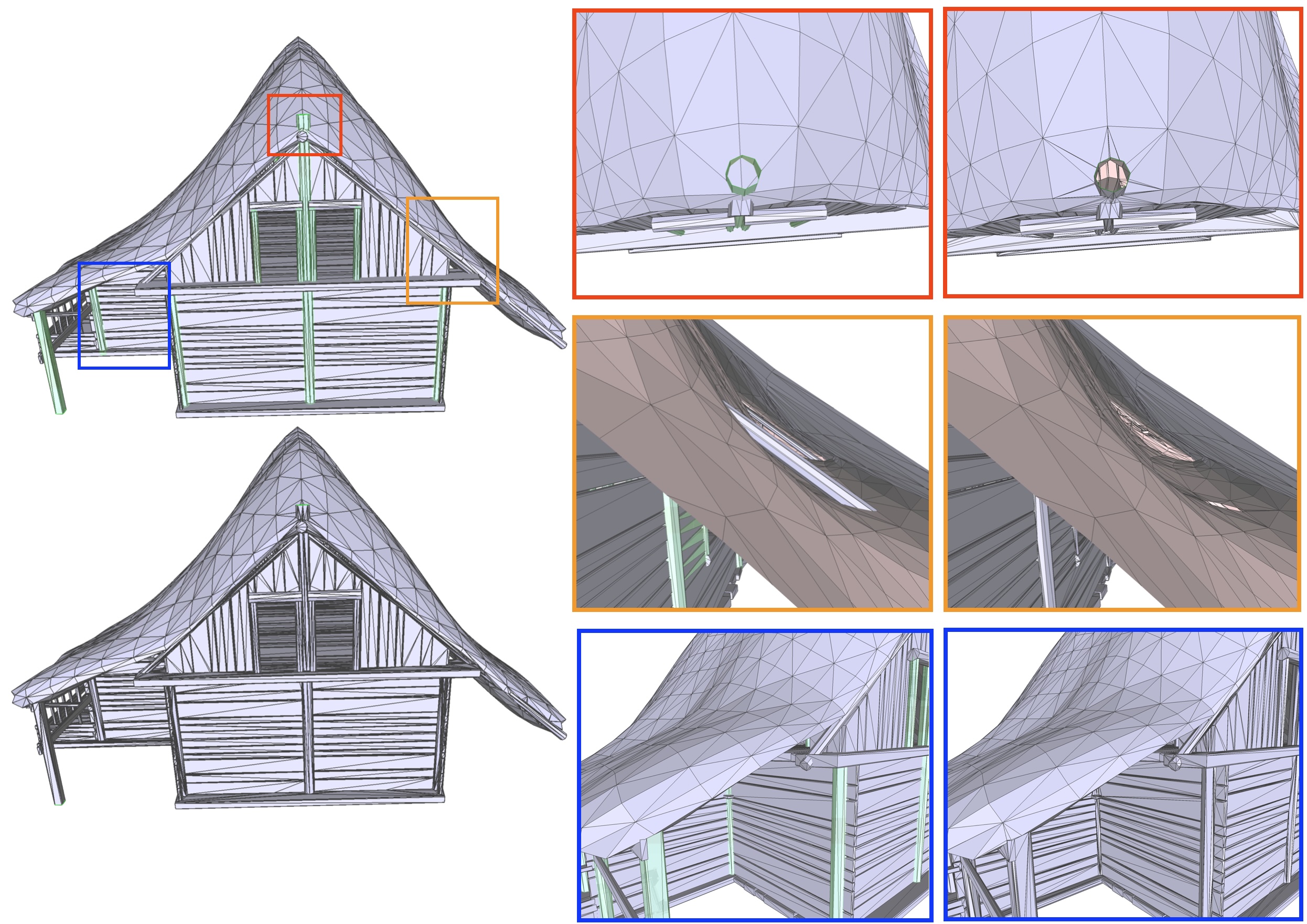}}
\caption{For damaged input models, when the normal vectors are correct, the outward offset with $0.1\%l$ can correct the mesh's minor gaps, self-intersections, and internal redundant faces. Green triangles indicate the border on the mesh. The red box show the fix of delete the error facet, because of the limitation of resolution with figure the boundary triangle mark with green is not obviously, this triangle only have the width of the offset distance. The input mesh has 6170 points, 9991 faces, 416 disconnected components, and 17808 pairs of self-intersecting faces. 10.8 percent  of the triangles are located on the boundary of the mesh. After performing offset calculations, the mesh has 20247 points, 41663 faces, and 6 disconnected components. After checking under rational numbers, 0 pairs of self-intersecting faces were found. After saving as an OBJ file with six significant digits for floating-point numbers and reopening it for checking, 947 pairs of self-intersecting faces were found, with 0.7 percent of the triangles located on the mesh boundary.}
\label{fig:showfix0}
\end{figure}
% \begin{figure}[t]
% \centering
% {\includegraphics[width=\linewidth]{fig/showfix1.jpg}}
% \caption{For damaged input models, when the normal vectors are correct, the outward offset with $0.1\%l$ can correct the mesh's minor gaps, self-intersections, and internal redundant faces. Green triangles indicate the border on the mesh. The input mesh has 5825 points, 9967 faces, 410 disconnected components, and 18760 pairs of self-intersecting faces. 2 percent of the triangles are located on the boundary of the mesh. After performing offset calculations, the mesh has 18653 points, 37858 faces, and 41 disconnected components. After checking under rational numbers, 0 pairs of self-intersecting faces were found. After saving as an OBJ file with six significant digits for floating-point numbers and reopening it for checking, 367 pairs of self-intersecting faces were found, with 0.4 percent of the triangles located on the mesh boundary.}
% \label{fig:showfix1}
% \end{figure}
Our approach can perform mesh repairs if the input mesh contains small gaps, self-intersections and internal redundant faces, etc. \prettyref{fig:showfix0} demonstrates the repair capability of our approach on such a case. Since our approach works nicely for small offset distances, it's particularly beneficial for repairing meshes while introducing minor changes to the original mesh. 

\subsection{Comparisons}
\begin{figure*}[h!]
\centering
\includegraphics[width=0.90\linewidth]{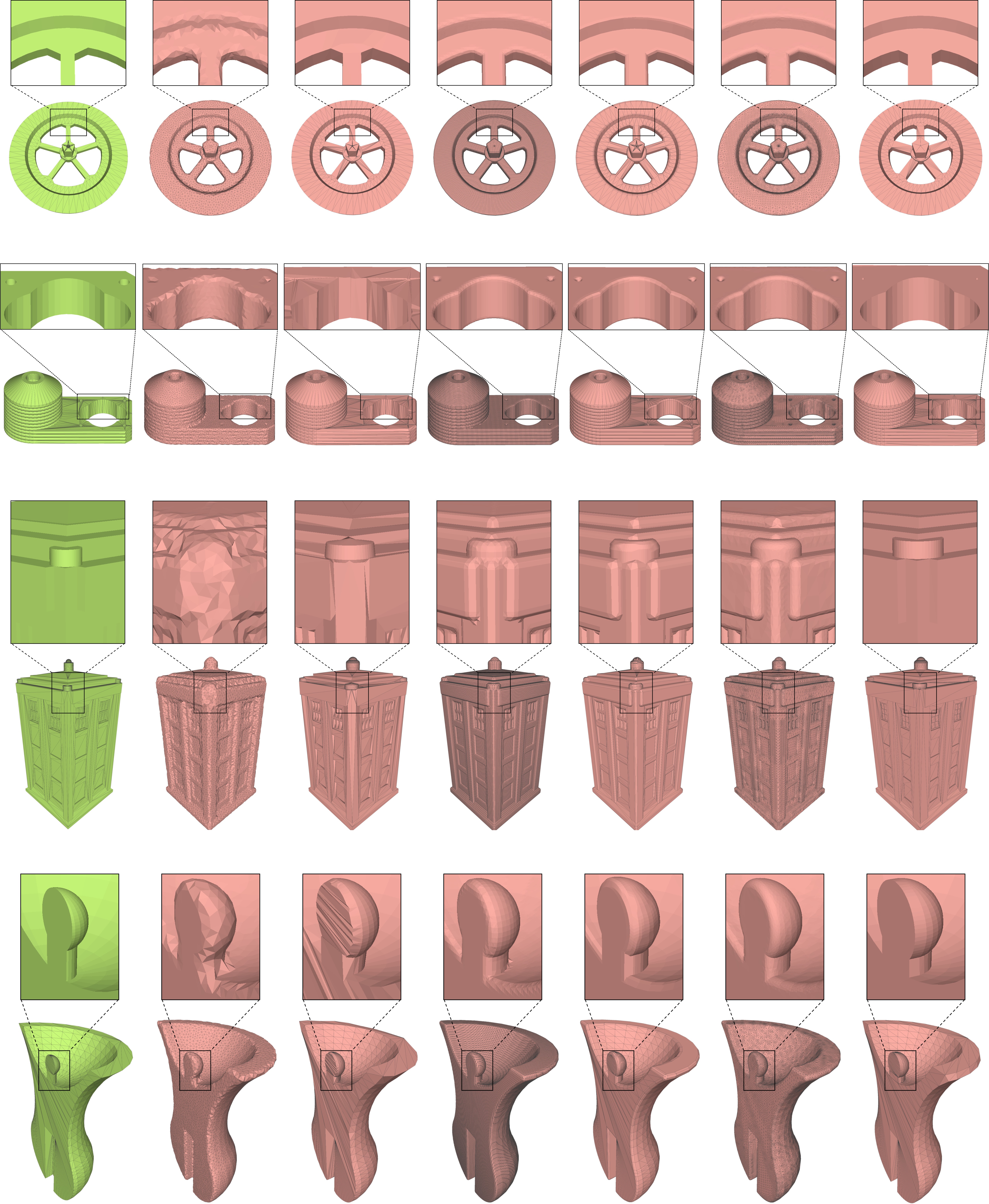}
 \footnotesize{
\put(-437,561){Input}
\put(-388,561){\citepalias{portaneri2022alpha} }
\put(-322,561){\cite{jung2004self}}
\put(-255,561){\cite{zhen2023}}
\put(-193,561){[\citetalias{cgal:eb-23b}]}
\put(-122,561){\cite{zint2023feature}}
\put(-40,561){Ours}
\put(-435,452){5354}
\put(-378,452){12894}
\put(-312,452){5710}
\put(-244,452){262360}
\put(-177,452){19838}
\put(-112,452){394470}
\put(-40,452){6468}
\put(-435,443){-}
\put(-378,443){2s}
\put(-312,443){925s}
\put(-244,443){84s}
\put(-177,443){1811s}
\put(-112,443){904s}
\put(-40,443){18s}
\put(-435,343){6240}
\put(-378,343){18724}
\put(-312,343){3244}
\put(-244,343){137928}
\put(-177,343){12402}
\put(-112,343){68560}
\put(-40,343){9878}
\put(-435,334){-}
\put(-378,334){6s}
\put(-312,334){1366s}
\put(-244,334){63s}
\put(-177,334){731s}
\put(-112,334){103s}
\put(-40,334){12s}
\put(-435,167){2456}
\put(-378,167){21942}
\put(-312,167){2726}
\put(-244,167){169712}
\put(-177,167){5600}
\put(-112,167){130578}
\put(-40,167){3120}
\put(-435,158){-}
\put(-378,158){8s}
\put(-312,158){1113s}
\put(-244,158){60s}
\put(-177,158){2892s}
\put(-112,158){197s}
\put(-40,158){12s}
\put(-435,-6){8642}
\put(-378,-6){13680}
\put(-312,-6){9276}
\put(-244,-6){238256}
\put(-177,-6){14352}
\put(-112,-6){270438}
\put(-40,-6){10524}
\put(-435,-15){-}
\put(-378,-15){9s}
\put(-312,-15){2133s}
\put(-244,-15){82s}
\put(-177,-15){1641s}
\put(-112,-15){152s}
\put(-40,-15){71s}
 }
\caption{\label{fig:comp}This figure shows a comparison of our method with other methods for outward offsets. It is clearly visible that our method has an advantage in restoring sharp features and achieves better detail restoration. The two value recorded in each sub-figure are generating facets number and running time.}
\end{figure*}

\begin{figure}[h!]
\centering
\includegraphics[width=\linewidth]{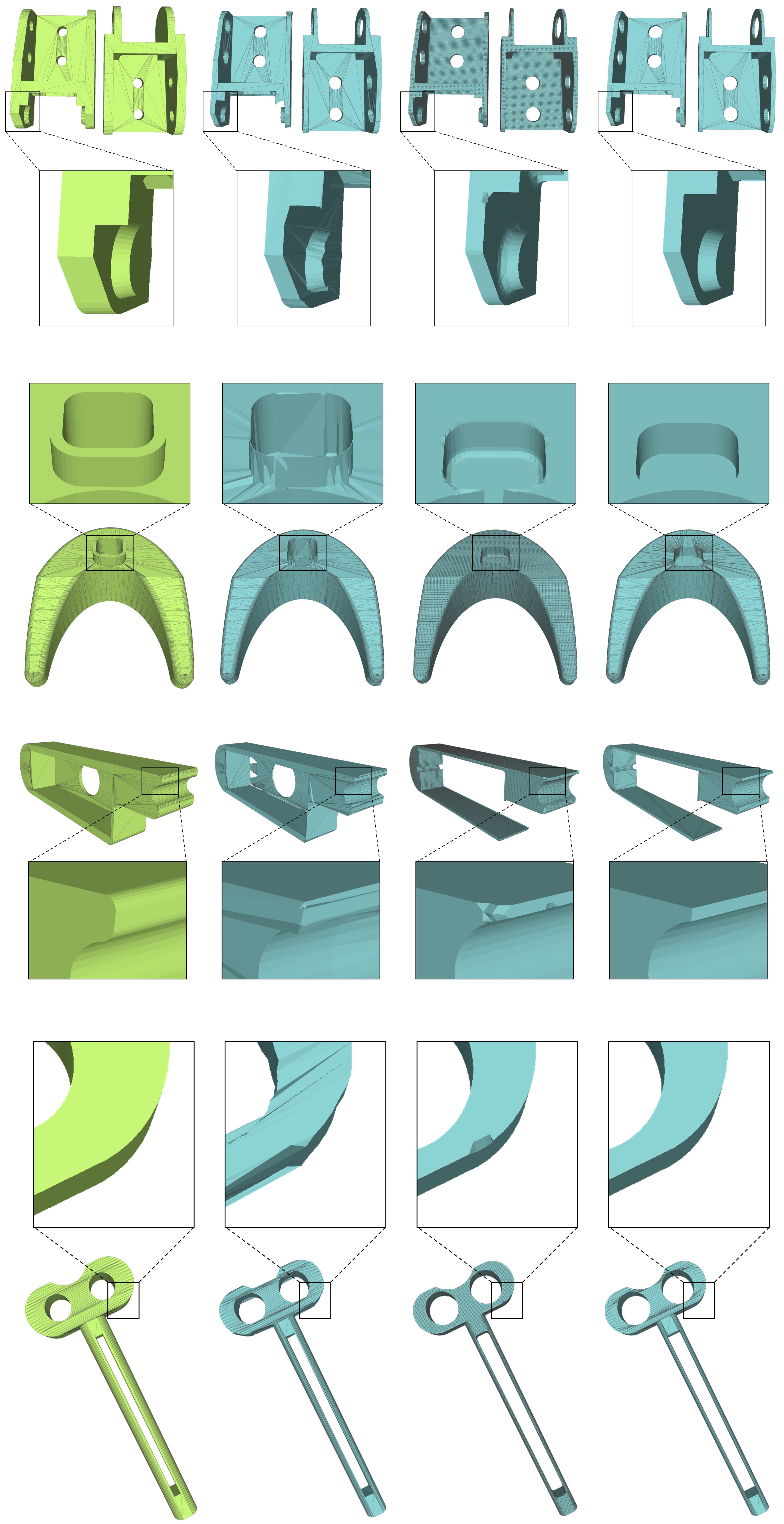}
\footnotesize{
\put(-220,476){Input}
\put(-170,476){\cite{jung2004self}}
\put(-110,476){\cite{zhen2023}}
\put(-35,476){Ours}
\put(-220,364){2682}
\put(-160,364){4420}
\put(-100,364){170296}
\put(-30,364){3565}
\put(-220,355){-}
\put(-160,355){1483s}
\put(-100,355){70s}
\put(-30,355){30s}
\put(-220,253){5016}
\put(-160,253){5890}
\put(-100,253){121256}
\put(-30,253){5732}
\put(-220,244){-}
\put(-160,244){958s}
\put(-100,244){63s}
\put(-30,244){20s}
\put(-220,161){908}
\put(-160,161){2598}
\put(-100,161){76072}
\put(-30,161){744}
\put(-220,152){-}
\put(-160,152){558s}
\put(-100,152){11s}
\put(-30,152){18s}
\put(-220,0){840}
\put(-160,0){3490}
\put(-100,0){42440}
\put(-30,0){2150}
\put(-220,-9){-}
\put(-160,-9){1745s}
\put(-100,-9){36s}
\put(-30,-9){16s}
}
\caption{\label{fig:comp2} This figure shows a comparison of our method with other methods for inward offsets. It is clearly visible that our method has an advantage in restoring sharp features and achieves better detail restoration. And \cite{jung2004self} can cause the error facet like show in the third row. The two value recorded in each sub-figure are generating facets number and running time.}
\end{figure}

\begin{figure}[h!]
\begin{tabular}{cc}
\includegraphics[width=\linewidth]{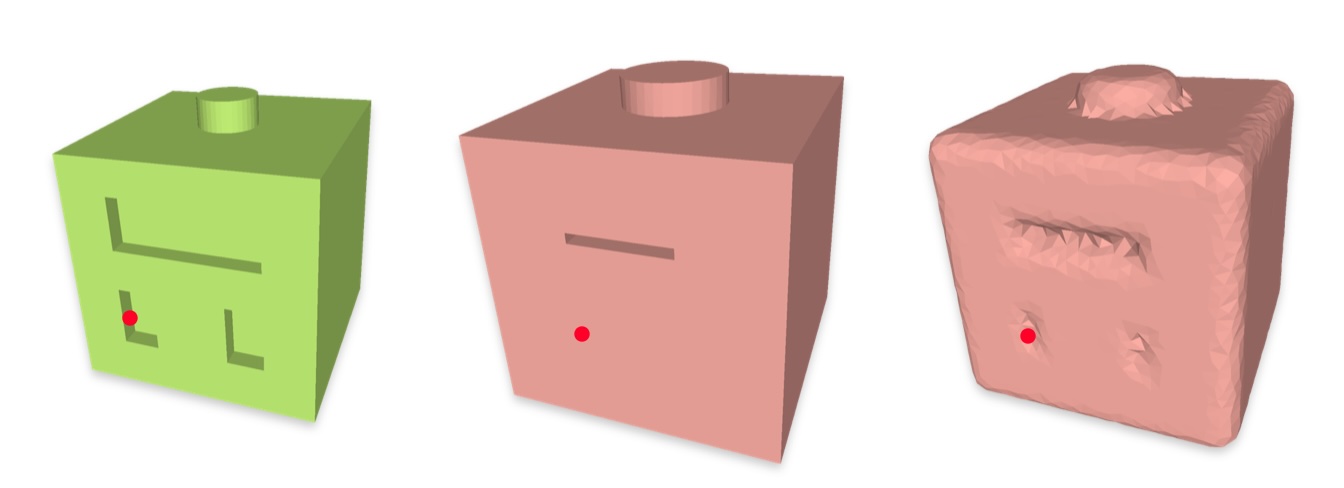}
\end{tabular}
\caption{\label{fig:zhibiaoyong} 
From left to right, the figures represent the input, the output of our method after a 5\%l offset, and the output of \citepalias{portaneri2022alpha}. The $D_{plane}(\MI, \MO)$ from the red point on the left model to the red point on the middle model is not as close to 5\%l as the $D_{plane}(\MI, \MO)$ from the left model to the right model.
}
\end{figure}

\begin{figure*}[h!]
\centering
\includegraphics[width=\linewidth]{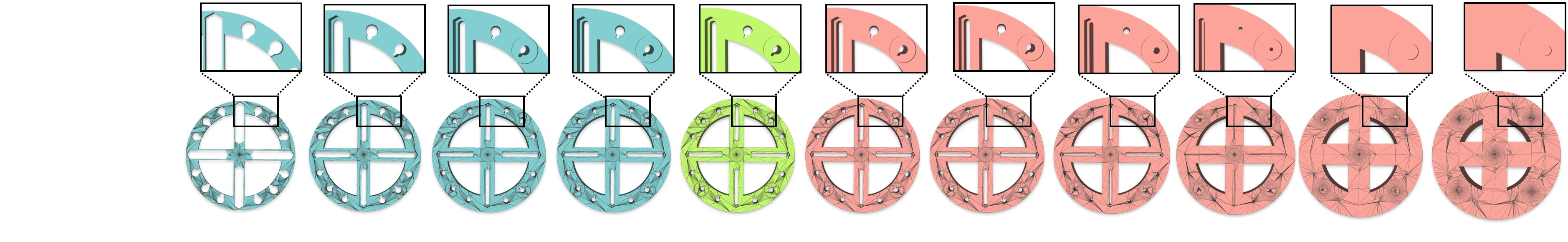}
\footnotesize{
\put(-489,27){Empty}
\put(-491,-7){$d=2\%$}
\put(-491,-14){$s=-1$}
\put(-441,-7){$d=1\%$}
\put(-441,-14){$s=-1$}
\put(-403,-7){$d=0.5\%$}
\put(-403,-14){$s=-1$}
\put(-364,-7){$d=0.1\%$}
\put(-364,-14){$s=-1$}
\put(-323,-7){$d=0.05\%$}
\put(-323,-14){$s=-1$}
\put(-279,-7){$input$}
\put(-241,-7){$d=0.05\%$}
\put(-241,-14){$s=1$}
\put(-200,-7){$d=0.1\%$}
\put(-200,-14){$s=1$}
\put(-159,-7){$d=0.5\%$}
\put(-159,-14){$s=1$}
\put(-118,-7){$d=1\%$}
\put(-118,-14){$s=1$}
\put(-78,-7){$d=2\%$}
\put(-78,-14){$s=1$}
\put(-34,-7){$d=3\%$}
\put(-34,-14){$s=1$}
}
\caption{\label{fig:varyd}show the mesh in different offset to generate a empty mesh.}
\end{figure*}

\begin{table*}[]
\scriptsize
\begin{tabular}{cccccccccccc}
d & s & method & \tiny{$D_{plane}({M_o,M_i})$} & \tiny{$D_{point}({M_o,M_i})$} & \tiny{$D_{angle}({M_o,M_i})$} & \tiny{$D_{plane}({M_i,M_o})$} & \tiny{$D_{point}({M_i,M_o})$} & \tiny{$D_{angle}({M_i,M_o})$} & SUC & FACE & TIME \\  \hline
$0.05\%l$ & -1 & \cite{jung2004self}  & 0.000132 & 0.000127 & 1.54531 & N/A & N/A & 2.18058 & 95 & \textbf{1938.6} & 281.5 \\
 &  & \cite{zhen2023}  & - & - & - & N/A & N/A & - & 0 & - & - \\
 &  & Ours & \textbf{0.000002} & \textbf{0.000006} & \textbf{0.193007} & N/A & N/A & \textbf{0.654405} & \textbf{100} & 2333.2 & \textbf{8.6} \\ \hline
$0.1\%l$ & -1 & \cite{jung2004self}  & 0.000265 & 0.000255 & 2.53419 & N/A & N/A & 3.4676 & 95 & \textbf{1955.3} & 203.9 \\
 &  & \cite{zhen2023}  & - & - & - & N/A & N/A & - & 0 & - & - \\
 &  & Ours & \textbf{0.000005} & \textbf{0.000012} & \textbf{0.265377} & N/A & N/A & \textbf{0.68033} & \textbf{100} & 2344.6 & \textbf{7.4} \\ \hline
$0.5\%l$ & -1 & \cite{jung2004self}  & 0.001382 & 0.001318 & 7.95833 & N/A & N/A & 9.77958 & 94 & \textbf{2010.2} & 235.4 \\
 &  & \cite{zhen2023}  & 0.000055 & \textbf{0.000011} & 2.89306 & N/A & N/A & 2.64754 & 93 & 643734 & 70.4 \\
 &  & Ours & \textbf{0.000052} & 0.000061 & \textbf{0.905081} & N/A & N/A & \textbf{1.51099} & \textbf{100} & 2576.2 & \textbf{15.6} \\ \hline
$1\%l$ & -1 & \cite{jung2004self}  & 0.003081 & 0.002937 & 12.3374 & N/A & N/A & 14.0914 & 88 & \textbf{1869.9} & 236.3 \\
 &  & \cite{zhen2023}  & 0.000219 & \textbf{0.000029} & 5.84421 & N/A & N/A & 5.08535 & \textbf{100} & 172471 & 46.8 \\
 &  & Ours & \textbf{0.000149} & 0.000116 & \textbf{2.10864} & N/A & N/A & \textbf{3.43261} & \textbf{100} & 2756 & \textbf{30.4} \\ \hline
$5\%l$ & -1 & \cite{jung2004self}  & 0.027254 & 0.026503 & 28.4349 & N/A & N/A & 24.1626 & 58 & 1238 & 539.9 \\
 &  & \cite{zhen2023}  & 0.005515 & 0.000722 & 13.4986 & N/A & N/A & 12.2177 & 95 & 8459 & \textbf{1.5} \\
 &  & Ours & \textbf{0.002249} & \textbf{0.000198} & \textbf{7.70095} & N/A & N/A & \textbf{7.84146} & \textbf{100} & \textbf{887} & 115.4 \\ \hline
$0.05\%l$ & 1 & $[$\citetalias{portaneri2022alpha}$]$ & 0.000131 & 0.00016 & 9.70743 & \textbf{0.000117} & \textbf{0.000101} & 9.54859 & \textbf{100} & 21239 & \textbf{6} \\
 &  & \cite{jung2004self}  & 0.000135 & 0.00013 & 1.35883 & 0.000296 & 0.000286 & 1.58863 & 95 & \textbf{1898.6} & 320.5 \\
 &  & \cite{zhen2023}  & - & - & - & - & - & - & 0 & - & - \\
 &  & \cite{zint2023feature} & 0.000419 & 0.00043 & 6.17889 & 0.001944 & 0.001733 & 6.00585 & 14 & 9334070 & 104.277 \\
 &  & $[$\citetalias{cgal:eb-23b}$]$ & 0.000711 & 0.000733 & 5.87534 & 0.000137 & 0.000151 & 5.83383 & 79 & 4967.2 & 425.1 \\
 &  & Ours & \textbf{0.000001} & \textbf{0.000006} & \textbf{0.432793} & 0.000243 & 0.000239 & \textbf{0.868208} & \textbf{100} & 2314.1 & 6.6 \\ \hline
$0.1\%l$ & 1 & $[$\citetalias{portaneri2022alpha}$]$ & 0.000098 & 0.000128 & 12.427 & 0.00038 & 0.000365 & 12.7267 & \textbf{100} & 17915.2 & \textbf{5.1} \\
 &  & \cite{jung2004self}  & 0.000274 & 0.000267 & 2.19128 & 0.000616 & 0.000605 & 2.54556 & 93 & \textbf{1860.5} & 209.2 \\
 &  & \cite{zhen2023}  & - & - & - & - & - & - & 0 & - & - \\
 &  & \cite{zint2023feature} & 0.000132 & 0.000147 & 2.54262 & 0.001364 & 0.001165 & 4.16945 & 38 & 706214 & 247.291 \\
 &  & $[$\citetalias{cgal:eb-23b}$]$ & 0.000388 & 0.000413 & 6.30141 & \textbf{0.00029} & \textbf{0.000281} & 6.44548 & 79 & 4965.8 & 404.5 \\
 &  & Ours & \textbf{0.000002} & \textbf{0.000007} & \textbf{0.474414} & 0.000495 & 0.000488 & \textbf{0.94634} & \textbf{100} & 2329.3 & 6.6 \\ \hline
$0.5\%l$ & 1 & $[$\citetalias{portaneri2022alpha}$]$ & 0.000179 & 0.000074 & 20.6634 & \textbf{0.002488} & \textbf{0.002444} & 20.8965 & \textbf{100} & 15416 & \textbf{4} \\
 &  & \cite{jung2004self}  & 0.001429 & 0.001351 & 6.43899 & 0.003207 & 0.003171 & 6.93395 & 88 & \textbf{1734.9} & 294.8 \\
 &  & \cite{zhen2023}  & 0.000137 & \textbf{0.00001} & 6.73193 & 0.002582 & 0.002547 & 4.12022 & 90 & 570871 & 82.6 \\
 &  & \cite{zint2023feature} & 0.000159 & 0.00004 & 2.19956 & 0.003191 & 0.003152 & 3.75481 & 77 & 180346 & 176.7 \\
 &  & $[$\citetalias{cgal:eb-23b}$]$ & 0.000346 & 0.00022 & 6.80817 & 0.002668 & 0.002638 & 6.97272 & 79 & 4858.7 & 422.9 \\
 &  & Ours & \textbf{0.000119} & 0.000054 & \textbf{1.11749} & 0.0026 & 0.002565 & \textbf{1.52496} & \textbf{100} & 2562.3 & 15.2 \\ \hline
$1\%l$ & 1 & $[$\citetalias{portaneri2022alpha}$]$ & 0.000575 & 0.000116 & 22.9292 & \textbf{0.005089} & \textbf{0.004997} & 21.6803 & \textbf{100} & 15900.4 & \textbf{3.4} \\
 &  & \cite{jung2004self}  & 0.002953 & 0.002674 & 9.54129 & 0.006582 & 0.006485 & 9.971 & 86 & \textbf{1761} & 438.2 \\
 &  & \cite{zhen2023}  & 0.000529 & 0.000037 & 8.53625 & 0.005337 & 0.005237 & 6.42665 & \textbf{100} & 126664 & 50.3 \\
 &  & \cite{zint2023feature} & 0.000427 & \textbf{0.000023} & 4.7115 & 0.006542 & 0.006432 & 6.10771 & 94 & 112318 & 160.5 \\
 &  & $[$\citetalias{cgal:eb-23b}$]$ & 0.000948 & 0.000476 & 7.85758 & 0.005553 & 0.005463 & 7.55089 & 79 & 4639.1 & 403.3 \\
 &  & Ours & \textbf{0.000303} & 0.000202 & \textbf{1.33165} & 0.005403 & 0.005302 & \textbf{1.79176} & \textbf{100} & 2950.9 & 42.8 \\ \hline
$5\%l$ & 1 & $[$\citetalias{portaneri2022alpha}$]$ & 0.007818 & \textbf{0.000074} & 19.9315 & \textbf{0.029172} & \textbf{0.028471} & 18.6799 & \textbf{100} & 20533.5 & 2.7 \\
 &  & \cite{jung2004self}  & 0.017843 & 0.013372 & 18.7543 & 0.037292 & 0.036438 & 18.0737 & 60 & \textbf{1319.8} & 780 \\
 &  & \cite{zhen2023}  & 0.008097 & 0.000626 & 14.0047 & 0.029311 & 0.028586 & 13.6277 & \textbf{100} & 4862.7 & \textbf{1.5} \\
 &  & \cite{zint2023feature} & 0.007315 & 0.000259 & 13.8445 & 0.03167 & 0.030529 & 13.8605 & 97 & 11977.9 & 45.6 \\
 &  & $[$\citetalias{cgal:eb-23b}$]$ & 0.009682 & 0.002429 & 14.7866 & 0.030825 & 0.029757 & 13.5377 & 79 & 3301.1 & 418.7 \\
 &  & Ours & \textbf{0.006067} & 0.003044 & \textbf{5.09495} & 0.030913 & 0.02971 & \textbf{5.11666} & 94 & 3062.4 & 266.1 \\ \hline
\end{tabular}
\caption{\label{tab:newcomp1} This table displays the distance metrics and angle metrics for all methods. To ensure fairness, if a method cannot complete the generation of half of the models in the dataset within the constrained time at a certain offset distance, we will not evaluate it. It can be observed that our method tends to satisfy the point-to-plane distance more compared to the Hausdorff distance. Moreover, methods that generate a larger number of faces have an advantage in satisfying the Hausdorff distance. The column of angle means the different dihedral angle of sampling points in feature line. It can be easy to see our method have a good performance in maintain feature. The last three columns display the number of models successfully completed by all methods (denoted as SUC), the number of faces generated (denoted as FACE), and the average time consumed (denoted as TIME). The time cost of six timeout model of ours are
$3602.65s, 40004.5s, 5539.63s, 4476.96s, 13355.2s, 3679.04s$.}
\end{table*}

% \usepackage{multirow}

To demonstrate the robustness and effectiveness of our approach, we compare against five state-of-the-art competing approaches, i.e. \citepalias{portaneri2022alpha}, \cite{zhen2023}, [\citetalias{cgal:eb-23b}], \cite{zint2023feature}, and \cite{jung2004self}, by batch processing the entire test dataset with varying offset distance settings, i.e. $d=0.05\%l$, $d=0.1\%l$, $d=0.5\%l$, $d=1\%l$ and $d=5\%l$ both inwardly (s = -1) and outwardly (s = 1). 
For outward offset, we compute the two way distances 
$H_{point}(\MO, \MI)$, $H_{point}(\MI, \MO)$, $H_{plane}(\MO, \MI)$, $H_{plane}(\MI, \MO))$, $D_{angle}(S_{\MO}, S_{\MI})$, and $D_{angle}(S_{\MI}, S_{\MO})$. However, for the inward offset surface, we evaluate the single way distance only, i.e. $H_{point}(\MO, \MI)$, $H_{plane}(\MO, \MI)$, and $D_{angle}(S_{\MO}, S_{\MI})$. The reason is that the inward offset surface may diminish as the offset distance gets large, as shown in \prettyref{fig:varyd}, which renders the computed distances of $H_{point}(\MI, \MO)$, $H_{plane}(\MI, \MO)$, and $D_{angle}(S_{\MI}, S_{\MO})$ not necessarily meaningful. 
We set a maximum execution time of one hour for each algorithm on each model. For processing time more than an hour, we automatically terminate the process and mark it as a failure. 
While we elaborate the detailed comparisons below, \prettyref{fig:feature1} and \prettyref{fig:comp} show visual comparisons for outward offset results generated by the various methods, \prettyref{fig:feature_line_in} and \prettyref{fig:comp2} illustrate the visual differences for inward offset meshes produced by the comparing approaches, and \prettyref{tab:newcomp1} demonstrates all the quantitative statistics. 

% \paragraph{Comparison with \citepalias{portaneri2022alpha}}  
% \begin{figure}[h!]
% \flushleft
% \setlength{\tabcolsep}{0pt}
% \begin{tabular}{cc}
% \includegraphics[width=\linewidth]{fig/zhibiaoyong.jpg}
% \end{tabular}
% \caption{\label{fig:zhibiao}From this figure, we can see that the first line from left to right is the input, the result based on the distance method, and our result.From $M_o$ to $M_i$, as shown in the orange line in the second row of the figure, our method $D_{plane}$ will have a better result indicator. From $M_o$ to $M_i$ in the second row, as shown in the red line in the figure, our results from $M_i$ to $M_o$ and distance-based methods have similar results. However, for the fourth row of the figure, our method, distance indicator $D_{plane}$, will show inferior performance compared to distance-based methods in this local area.The fundamental reason is that our method aims to preserve sharp features, resulting in larger offset distances in some local regions. This makes it easier for these regions to be fused than distance-based methods, which can lead to a lower $D_{plane}$ metric for $M_i$ to $M_o$ in these regions compared to distance-based methods.}
% \end{figure}

\citepalias{portaneri2022alpha} can robustly generate a watertight and orientable surface triangle mesh from an arbitrary 3D geometry input, but cannot compute inward offsets. Therefore, as shown in \prettyref{tab:newcomp1}, we compare with it for cases only when $s=1$. It has two parameters, i.e. an alpha to determine the size of untraversable cavities when refining and carving a 3D Delaunay triangulation and an offset distance to control how far of the output mesh vertices from the input. 
We compare our algorithm with \citepalias{portaneri2022alpha}'s CGAL implementation, set its offset distance using $d$. Its alpha parameter plays a critical role in determining the output mesh's geometry approximation of the input and the approach's computational speed. Using the same setting in the experiment section of \citepalias{portaneri2022alpha}, we set alpha = 100 for all the test in this paper.
For the point-to-point distance $D_{point}$, while \citepalias{portaneri2022alpha} performs excellently when measuring the distance from $\MI$ to $\MO$, as shown in the $D_{point}(\MI, \MO)$ column of \prettyref{tab:newcomp1}, our approach has generally better accuracy when computing the distance from $\MO$ to $\MI$. Similar pattern happens for the point-to-plane distance $D_{plane}$ that our approach consistently achieves the smallest error for $D_{plane}(\MO, \MI)$ for all the offset distances, but \citepalias{portaneri2022alpha} has the best accuracy for $D_{plane}(\MI, \MO)$. While this seems unexpected since $D_{plane}$ is designed for measuring the mitering offset distances, our approach may remove tiny dents as show in \prettyref{fig:zhibiaoyong} that can lead to large $D_{plane}(\MI, \MO)$ and $D_{point}(\MI, \MO)$. In fact, our approach captures the sharp feature of the input excellently, achieving the best feature preservation among all the approaches for all the tested offset distances, as shown by the $D_{angle}(\MI, \MO)$ and $D_{angle}(\MO, \MI)$ columns of \prettyref{tab:newcomp1}, and \prettyref{fig:feature1}.

\paragraph{Comparison with \cite{zhen2023}} We compare our algorithm with \cite{zhen2023} by directly running their provided online executable program. While \cite{zhen2023} can robustly handle general inputs, relying on voxel discretization of the unsigned distance function and maching-cube-like linear approximation of the iso-surface, \cite{zhen2023} can generate results only when the discretization is coarse as shown by the $D_{point}(\MO, \MI)$ for $d = 0.5\%l, s = -1$, $d = 1\%l, s = -1$ and $d = 0.5\%l, s = 1$. However, the provided implementation would incur memory and computational issues for small offset distances, which is denoted as ``$-$'' in \prettyref{tab:newcomp1} for program crashes. Although introducing the feature preserving EMC33 scheme, their results tend to produce creases, leading to noisy feature curves as illustrated in \prettyref{fig:feature1} and \prettyref{fig:feature_line_in}.

\paragraph{Comparison with \cite{zint2023feature}}We compare our algorithm with \cite{zint2023feature} by running directly the source code provided by the authors. Their method also generates the mesh through implicit iso-surface extraction, similar to \cite{zhen2023}. By using octree, it can improve performance at small offset distances. The results of this algorithm are controlled by three main parameters: the minimum octree depth (d0), the maximum octree depth (d1), and the octree depth for resolving non-manifold vertices (d2). We use the default setting, i.e. d0 = 0, d1 = 10, and d2 = 12. Due to the usage of dual contouring for iso-surface extraction, this approach preserves feature lines very well, only second to our algorithm as show in the $D_{angle}(\MI, \MO)$ and $D_{angle}(\MO, \MI)$ columns of \prettyref{tab:newcomp1} and as show in \prettyref{fig:feature1}. It can get the best or close to the best result in $D_{point}(\MO, \MI)$ metric. For all the tested offset distances, \cite{zint2023feature} produces results with consistently large dense meshes as compared to most of the other approaches. Especially as the offset distance gets small, the face numbers of \(\MO\) can reach to millions of triangles, e.g. over 9M for $d = 0.05\%l$. Furthermore, their provided program never successfully processes the entire dataset for the different batch tests. Due to unknown issues, the code provided by \cite{zint2023feature} cannot handle inward offsets, therefore, we only perform the outward offset experiments.

\paragraph{Comparison with [\citetalias{cgal:eb-23b}]} 
This method employs the computational results of the Minkowski sum of a sphere (represented as a polyhedron) and $M_i$. It has a corresponding function available in version 5.6 of the CGAL library, released in 2023, and exclusively supports outward offsets. Instead of related to offset distances, the computational performance of this method is unstable and heavily depends on the model's concavity and convexity, with longer computation times observed for models with more concave features. If a model strictly adheres to the definition of a convex hull, the computational efficiency is high. It does not support meshes with self-intersections, open boundaries, etc. The evaluated measurements for $D_{point}$ and $D_{plane}$ for this approach do not exhibit particularly good results. Its output is represented by planar polygons, which, when converted into a triangular mesh, shows a lower number of faces compared to those obtained by the implicit iso-surface extraction methods. This approach preserves feature lines well, as shown in \prettyref{fig:feature1}, but does not maintain dihedral angles well, as indicated in the $D_{angle}(\MO, \MI)$ and $D_{angle}(\MI, \MO)$ columns of \prettyref{tab:newcomp1}.

\paragraph{Comparison with \cite{jung2004self}} 
Using C++ and CGAL, we implemented the algorithm presented in \cite{jung2004self}. This method achieves the offset surface by first directly moving vertices of $M_i$ following their normal directions, and then calculating a convex hull of all of the moving points. However, their approach can lead to missing or erroneous faces as shown in the second column of \prettyref{fig:comp2}. Moreover, in scenarios with complex and many self-intersections, this approach can have a significant increase in computational time as the offset distance increases, whereas our method does not show such a significant rise as shown in the \prettyref{tab:newcomp1} and \prettyref{fig:timeown} when offset distance become large like $d >= 1\%l$. For many cases, relying solely on the normal direction can easily produce a large number of creases, thus generating many redundant features as shown in \prettyref{fig:feature1} and \prettyref{fig:feature_line_in}. Additionally, this results in particularly poor $D_{angle}$ values. And its $D_{point}$ and $D_{plane}$ values often among the worst over all algorithm as shown in \prettyref{tab:newcomp1}. 

To summarize, our method can excellently capture sharp features of the inputs and has the best $D_{plane}(\MO,\MI)$, $D_{angle}(\MO, \MI)$, and $D_{angle}(\MI, \MO)$ among all methods because our approach focuses more on the mitered offset effect. Compared to the competing approaches, our method runs fast at small offset distances, and due to the presence of acceleration strategies, it still performs reasonably well as the offset distance increases.

\subsection{Timing.} 
% \begin{figure}
%     \centering
%     \includegraphics[width=0.9\columnwidth]{fig/bingtu.jpg}
%     \caption{\label{fig:time-plot}This figure shows the execution time of each part of our algorithm.}
% \end{figure}
% \begin{figure}
%     \centering
%     \includegraphics[width=1.0\columnwidth]{fig/quxian.jpg}
%     \tiny{
%     \put(-65,127.5){\cite{jung2004self}}
%      \put(-65,121){\cite{zhen2023}}
%      \put(-65,115){\cite{zint2023feature} }
%      \put(-65,108.5){$[$\citetalias{cgal:eb-23b}$]$ }
%      \put(-65,102){$[$\citetalias{portaneri2022alpha}$]$ }
%      \put(-65,95.5){Ours}
%     }
%     \caption{\label{fig:time-plot-comp}This figure shows the different method timing cost with 5 representative models. In 0.25\%l offset distance, \cite{zint2023feature} and \cite{zhen2023} can not success generate the offset mesh. }
% \end{figure}

\begin{figure*}[h!]
\flushleft
\setlength{\tabcolsep}{0pt}
\begin{tabular}{cc}
\includegraphics[width=\linewidth]{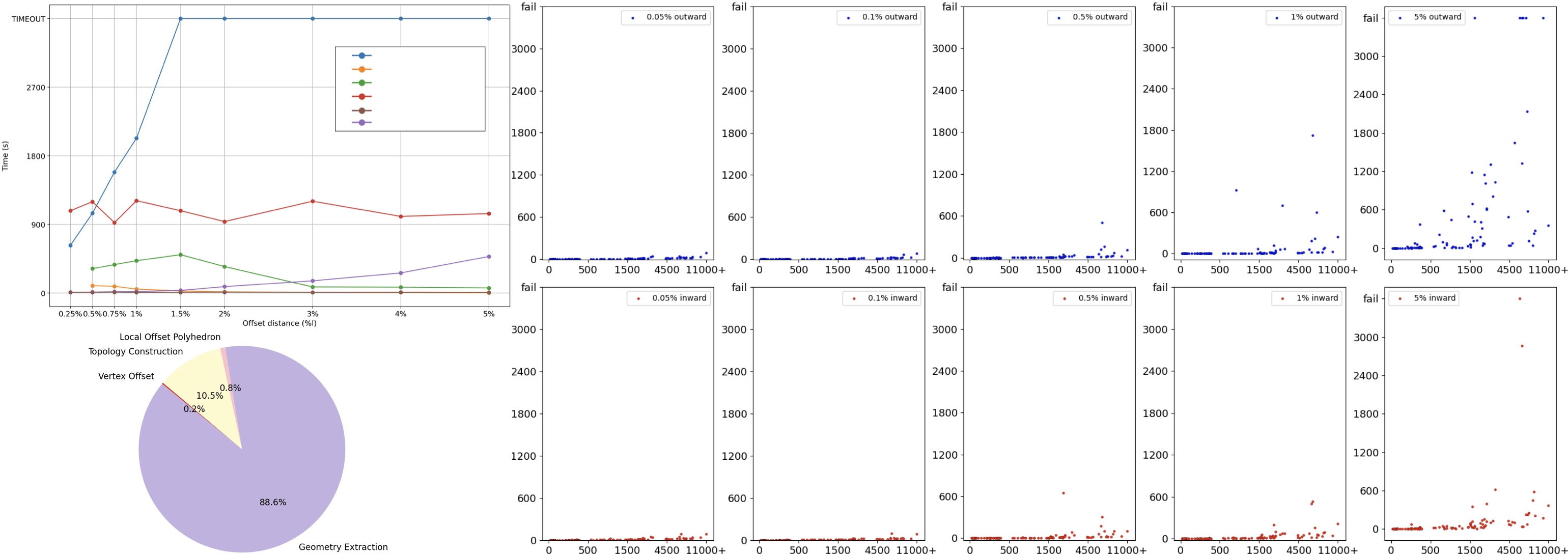}
\fontsize{4pt}{4pt}\selectfont{
    \put(-385,162){\cite{jung2004self}}
     \put(-385,157){\cite{zhen2023}}
     \put(-385,152.5){\cite{zint2023feature} }
     \put(-385,148){$[$\citetalias{cgal:eb-23b}$]$ }
     \put(-385,143.5){$[$\citetalias{portaneri2022alpha}$]$ }
     \put(-385,139){Ours}
    }
\end{tabular}
\caption{\label{fig:timeown}
Top-left corner of this figure shows the different method timing cost with 5 representative models. In 0.25\%l offset distance, \cite{zint2023feature} and \cite{zhen2023} can not success generate the offset mesh.
Bottom-left corner of this figure the execution time of each part of our algorithm.
The right part shows input triangle number VS time cost.}
\end{figure*}

As shown in the $T$ column of \prettyref{tab:newcomp1}, averaged over the entire tested dataset, our approach achieves the fastest computational speed for inward offset generation when offset distance $\leq 1\%$. For outward offset generation, while \cite{portaneri2022alpha} is consistently among the fastest approach, as the offset distance increases, ours changes from being comparable to \cite{portaneri2022alpha} at $0.05\%$ and $0.1\%$ and is more than 30 times faster compared to \cite{jung2004self}, \cite{zint2023feature}, and \citetalias{cgal:eb-23b}, more than 10 times faster compared to \cite{jung2004self}, \cite{zhen2023}, \cite{zint2023feature}, and \citetalias{cgal:eb-23b} when $d = 0.5\%$, about 10 times faster compared to \cite{jung2004self}, and \citetalias{cgal:eb-23b} when $d = 1\%$, and only 2-3 times faster than \cite{jung2004self}, and \citetalias{cgal:eb-23b} when $d = 5\%$.
Top-left corner of \prettyref{fig:timeown} further compares the competing approaches with ours on five selected representative models by varying the offset distances from $d=0.25\%$ to $5\%$. It's clear that our approach ranks among the fastest one when offset distance is small and the computation slows down for large offset distances. The reason behind is due to the increased amount of self-intersection computations of our approach when $d$ increases. 
Bottom-left corner of \prettyref{fig:timeown} further verifies that our computational bottleneck lies in the intersection resolving operations for models with excessive details. This step accounts $88.6$\% of the total computing time. Right half of \prettyref{fig:timeown} plots the detailed timing for each model by our approach for the varying offset distances.

% . Although with the speedup of the early filtering strategy, Our computational bottleneck still lies in the Boolean operations during the intersection resolving, especially for models with excessive details. This  step accounts $98.4$\% of the total computing time, i.e. $33.5\%$ by the early filtering and $64.9\%$ by the geometry extraction. 

% \begin{figure}[h!]
% \flushleft
% \setlength{\tabcolsep}{0pt}
% \begin{tabular}{cc}
% \includegraphics[width=0.99\linewidth]{fig/Figure_1direct.jpg}

% \end{tabular}
% \caption{\label{fig:timedirect}input facet VS time cost of  \cite{jung2004self}}
% \end{figure}

% \begin{figure}[h!]
% \flushleft
% \setlength{\tabcolsep}{0pt}
% \begin{tabular}{cc}
% \includegraphics[width=0.99\linewidth]{fig/Figure_1mc.jpg}
% \end{tabular}
% \caption{\label{fig:timemc}input facet VS time cost of \cite{zhen2023}}
% \end{figure}

% \begin{figure}[h!]
% \flushleft
% \setlength{\tabcolsep}{0pt}
% \begin{tabular}{cc}
% \includegraphics[width=0.99\linewidth]{fig/voxcel.jpg}
% \end{tabular}
% \caption{\label{fig:timevoxel}input facet VS time cost of \cite{zint2023feature}}
% \end{figure}

% \begin{figure}[h!]
% \flushleft
% \setlength{\tabcolsep}{0pt}
% \begin{tabular}{cc}
% \includegraphics[width=0.99\linewidth]{fig/mink.jpg}
% \end{tabular}
% \caption{\label{fig:timemink}input facet VS time cost of  Minkowski sum\cite{cgal:eb-23b}}
% \end{figure}

% \begin{figure}[h!]
% \flushleft
% \setlength{\tabcolsep}{0pt}
% \begin{tabular}{cc}
% \includegraphics[width=0.99\linewidth]{fig/alpha200.jpg}
% \end{tabular}
% \caption{\label{fig:timealpha200}input facet VS time cost of Alpha wrapping \cite{portaneri2022alpha} with $\alpha = 200$.}
% \end{figure}

% \begin{figure}[h!]
% \flushleft
% \setlength{\tabcolsep}{0pt}
% \begin{tabular}{cc}
% \includegraphics[width=0.99\linewidth]{fig/alpha20.jpg}
% \end{tabular}
% \caption{\label{fig:timealpha20}input facet VS time cost of Alpha wrapping \cite{portaneri2022alpha} with $\alpha = 20$.}
% \end{figure}

%% file: 05-conclusion.tex
\section{Conclusion}
We propose a new and robust approach to generate an offset mesh for a 3D input with an arbitrary topology and geometry complexities. 
Our method ensures the output with several nice properties, such as feature preservation, similar number of triangles with the input, free of self-intersections and degenerate elements. Our approach also support user-specified non-uniform offset distances. We anticipate that, with all these advantages combined, our method makes a significant advancement in the geometry processing related field.

\paragraph{Discussion and Limitations:}Several aspects worth to investigate to further improve the current approach. First, the generalized winding number may be computed wrongly for points very close to the mesh surface, which is an issue could be mitigated through the combination of ray-intersection checks. Second, to introduce as few as possible approximations to the offset mesh generation, we only consider conservative speedup strategies in the current version, where for applications in rendering and animation, we may further improve the performance by designing accelerations with a controllable tolerance to the geometric distance. 
% \begin{figure}[h!]
% \centering
% \includegraphics[width=\linewidth]{fig/zhezhou.jpg}

% \caption{\label{fig:zhezhou}show the ours method in some case which can expand the pleats.}
% \end{figure}